\documentclass{PoS2}
\usepackage[authoryear,square]{natbib}
\bibpunct{(}{)}{;}{a}{}{,}

\usepackage{graphicx,epstopdf}
\newcommand{\gtsim}{\mbox{{\raisebox{-0.4ex}{$\stackrel{>}{{\scriptstyle\sim}}$}}}}

\newcommand{\Msolar}{M$_{\odot}$}
\newcommand{\Msolaryr}{M$_{\odot}$~yr$^{-1}$}

\title{The star-formation history of the Universe with the SKA}

\ShortTitle{The star-formation history of the Universe}

\author{
{Matt J.~Jarvis}$^{1,2}$, 
Nick Seymour$^3$\thanks{Speaker}, Jose Afonso$^{4,5}$, Philip
Best$^6$, Rob Beswick$^7$, Ian Heywood$^{8,9}$, Minh Huynh$^{10}$, Eric
Murphy$^{11}$, Isabella Prandoni$^{12}$, Eva Schinnerer$^{13}$, Chris Simpson$^{14}$, Mattia
Vaccari$^2$, Sarah White$^1$
\\ 
$^1$Astrophysics, University of Oxford, Keble Road, Oxford, OX1
        3RH, UK; $^2$Physics Department, University of the Western
        Cape, Bellville 7535, South Africa; 
$^3$International Centre for Radio Astronomy Research, Curtin
University, Perth, Australia; 
$^4$ Instituto de Astrof\'{i}sica e Ci\^{e}ncias do Espa\c co, Universidade de Lisboa, OAL, Tapada da Ajuda, PT1349-018 Lisboa, Portugal;
$^5$ Departamento de F\'{i}sica, Faculdade de Ci\^{e}ncias, Universidade de Lisboa, Edif\'{i}cio C8, Campo Grande, PT1749-016 Lisbon, Portugal; 
$^6$ Institute for Astronomy, University of Edinburgh, Royal
Observatory, Blackford Hill, Edinburgh EH9 3HJ, UK;
$^{7}$Jodrell Bank Centre for Astrophysics, Alan Turing Building,
University of Manchester, Oxford Road, Manchester, M13 9PL, UK; 
$^{8}$CSIRO Astronomy \& Space Science, P.O. Box 76, Epping, NSW 1710, Australia
$^9$RATT, Department of Physics and Electronics, Rhodes University, P.O. Box 94, Grahamstown, South Africa
$^{10}$International Centre for Radio
Astronomy Research, University of Western Australia, Perth, Australia;
$^{11}$IPAC, Caltech, MC 220-6, Pasadena CA, 91125, USA; 
$^{12}$INAF-IRA Bologna, Via Gobetti 101, I-40129 Bologna, Italy;
$^{13}$Max-Planck-Institut fu\"ur Astronomie, K\"onigstuhl 17, 69117 Heidelberg, Germany;
$^{14}$Astrophysics Research Institute, Liverpool John Moores University, ic2 Building, 146 Brownlow Hill, Liverpool L3 5RF
\\
E-mail: \email{matt.jarvis@astro.ox.ac.uk}
}

\abstract{Radio wavelengths offer the unique possibility of tracing
  the total star-formation rate in galaxies, both obscured and
  unobscured. As such, they may provide the most robust measurement of
the star-formation history of the Universe. In this chapter we
highlight the constraints that the SKA can place on the evolution of
the star-formation history of the Universe, the survey area required
to overcome sample variance, the spatial resolution requirements, along with the multi-wavelength
ancillary data that will play a major role in maximising the scientific
promise of the SKA. The required combination of depth and resolution
means that a survey to  trace the star formation in the Universe
should be carried out with a facility that has a resolution of at
least $\sim 0.5$\,arcsec, with high
sensitivity at $< 1$~GHz.
We also suggest a strategy that will enable new
parameter space to be explored as the SKA expands over the coming decade.}

\FullConference{
Advancing Astrophysics with the Square Kilometre Array\\
June 8-13, 2014\\
Giardini Naxos, Italy}

\begin{document}

\section{Introduction}

Gaining a full understanding of the formation and evolution of galaxies
relies on our ability to trace the build-up of stellar mass over the
history of the Universe. Therefore we are required to obtain
observations that allow us to measure both ongoing star-formation
activity and the stellar populations that are already in place. The
older stars, which contain the bulk of the stellar mass, emit the majority of their radiation towards the near-infrared
part of the electromagnetic spectrum. Ongoing surveys with the
Visible-Infrared Survey Telescope for Astronomy (VISTA),   
the {\em Spitzer Space Telescope} and the {\em Hubble Space Telescope}
(HST) are adept at tracing such emission
to very high redshifts. However, tracing the current star-formation
activity in distant galaxies is a much more difficult problem. This is
because the hot young stars are very blue and consequently are much more
susceptible to obscuration by dust, making ultra-violet and optical
surveys incomplete \citep[for a review see][]{Madau&Dickinson2014}. Indeed, from measurements of the integrated
optical and infrared background radiation we know that around 50 per
cent of the light from stellar processes is obscured by dust \citep[e.g.][]{Takeuchi2005,Dole2006}.

This property has motivated a long line of telescopes with the aim of
detecting the reprocessed dust emission from these young stars at
far-infrared wavelengths. The most recent of these, the {\em Herschel
  Space Observatory}, has provided a wealth of new information on the
total star-formation rate (SFR) density in the Universe and how this evolves
\citep[e.g.][]{Vaccari2010,Dye2010,Lapi2011,Burgarella2013,Gruppioni2013,Magnelli2013}. However, even {\em Herschel} does not have the ability to
track the star formation in galaxies to faint levels at high
redshift, and the differences in dust temperature may also lead to
systematic biases \citep[e.g.][]{Smith2014}. The relatively poor
spatial resolution 
means that imaging surveys with {\em Herschel} are confusion-noise
dominated, rather than instrumental or sky-background
dominated. Other types of survey such as those targeting galaxies with
emission-lines excited by the young stars have also been
successful, but are generally limited to either expensive follow-up spectroscopy
of known galaxies \citep[e.g.][]{Erb2003,Gilbank2010}, or narrow-band imaging campaigns that are limited in the volume that they
can survey, due to the width of the specific filter
\citep[e.g.][]{Sobral2012,Drake2013}. Both of these methods are also
affected by dust obscuration, and multiple lines of the same atomic
species are required to obtain an estimate of the dust extinction, and
thus obtain an accurate star-formation rate (SFR). Therefore, we are currently unable to obtain a clear view
of the total star formation occurring over cosmic time, what galaxies
this occurs in and when, and also in what environments such
activity is stimulated or truncated.

As we move towards the next generation of deep radio continuum surveys, the
dominant radio source population will no longer be active galactic
nuclei (AGN), but
star-forming galaxies \citep[e.g.][]{Cram1998,Haarsma2000,Afonso2005,Seymour2008, Padovani2009, McAlpine2013},
although radio-quiet AGN may still make a significant contribution
\citep[e.g.][]{JarvisRawlings2004, Simpson2006, Smolcic2009,
  Bonzini2013,White2014}.
 The radio emission from these star-forming
galaxies is predominantly in the form of synchrotron emission from
relativistic electrons accelerated in supernova remnants, and
free-free emission from H{\sc ii} regions \citep[for a review see][]{Condon1992}. Both of these emission
processes are linked to stars of masses $M\, \gtsim$~8~\Msolar~that end
in core-collapse supernovae, and dominate the ionisation of
H{\sc ii} regions. Thus, it is not surprising that radio continuum emission, 
where dust obscuration is no longer an issue, has been used to infer
the SFRs of galaxies. However, current
observations at high redshift are limited to stacking experiments, where only average
properties of galaxies selected at other wavelengths are determined
\citep[e.g.][]{Karim2011,Zwart2014}. This constraint will be overcome
with the vast sensitivity of the SKA.


Therefore, as we move into the SKA era, using the radio continuum emission to
trace the star-formation history of the Universe will potentially
provide us with the first unbiased view of star formation using a
single waveband. Such surveys will happen on the same time frame as
other major imaging facilities and multi-object spectrographs on 8-m
class telescopes (see Section~\ref{sec:multiwavelength}). The
combination of these major facilities will enable us to investigate
galaxy evolution from the perspective of both massive statistical
studies, coupled with detailed studies of well-selected samples,
focusing on the role of redshift, galaxy mass, environment and
feedback from both supernovae and active galactic nuclei. The
SKA will contribute to all of these types of study. In this chapter
we provide an overview of the statistical power of the SKA in
determining the history of star formation in the Universe and how
this may depend on galaxy mass and environment. 

\section{Assumptions}

In what follows we will assume that we have both the ability to
disentangle star-formation from AGN emission, and 
photometric redshifts with an uncertainty of $\Delta z/(1+z) \sim
0.05$ up to $z\sim 6$, based on current optical and
near-infrared surveys \citep[e.g.][]{Jarvis2013}. We note that such a
precision for emission-line objects is difficult but feasible over the
coming decade.

We also base our estimates of the star-formation history on the
luminosity functions that underpin the semi-empirical extragalactic
sky simulations of \cite{Wilman2008, Wilman2010}. These simulations
continue to provide a very good description of the latest
source counts from various deep field surveys with the JVLA
\citep[e.g.][]{Condon2012,Vernstrom2014}. Although modifications may be required
to accurately reproduce the most recent results from e.g. {\em Herschel},
the general trends and evolution prescribed are relatively well-matched to our current understanding, and the extrapolations to
flux-density levels yet to be reached in the radio band are
constrained by observations at a range of other wavelengths. For full
details see \cite{Wilman2010}.

The receivers being considered for both SKA1-SUR and SKA1-MID means that any observations will cover a large bandwidth of around
$1$~GHz, however, for ease of comparison with previous work, we
adopt a single frequency. However, see Section~\ref{sec:SKA} for a
discussion of the impact of this assumption.

SFRs derived from radio observations
have predominantly been calibrated to the integrated far-infrared
emission, which is one of the most accurate and unbiased tracers of
star-formation in galaxies, due to the optically thin nature of the
dust to far-infrared emission \citep{deJong1985,Appleton2004, Ivison2010b,Jarvis2010, Bourne2011}. In this chapter we use the relation
between star-formation rate and radio luminosity as provided by \cite{Yun2001}, although we note that similar results are
obtained if we use the relation of \cite{Bell2003}, to investigate how the SKA can contribute to this field.

\section{The evolution of radio luminosity function of star-forming
  galaxies}\label{sec:rlf}

The most straightforward experiment to trace the
star-formation history of the Universe is to measure the evolution of
the radio luminosity function of star-forming galaxies \citep[e.g.][]{Hopkins2004,Smolcic2009sf}. We can then use the relation
between radio luminosity and star-formation rate derived by several
authors \citep[e.g.][]{Condon1992,Yun2001,Bell2003} to estimate the
total star-formation rate in the galaxy.

This 
requires several key measurements: 1) the radio flux density, 2) the redshift of the source, 3)
the fraction of radio emission that is due to
star formation, rather than from an AGN. The
first of these is obviously measured directly from the radio continuum
emission, however 2) and 3) are more problematic. We discuss 2) in
Section~\ref{sec:multiwavelength} and defer details of 3) to
separate chapters \citep[see ][]{McAlpine2014,Makhatini2014}.
 We note that
the large bandwidth will also allow in-band spectral index
measurements \citep[see e.g.][]{Rau2014}, thus removing a source of uncertainty in measuring a
monochromatic rest-frame luminosity.

To address the evolution of star-forming galaxies from radio surveys,
one also needs a tiered survey strategy whereby enough volume is
sampled at each cosmic epoch of interest in order to overcome sample
variance and gain a representative view of the Universe, from the
sparsest voids through to the densest clusters. We therefore consider
three tiers that we believe to be representative of the survey strategy
that could be conducted with the SKA in phase 1.

\begin{figure}
\includegraphics[width=7.5cm]{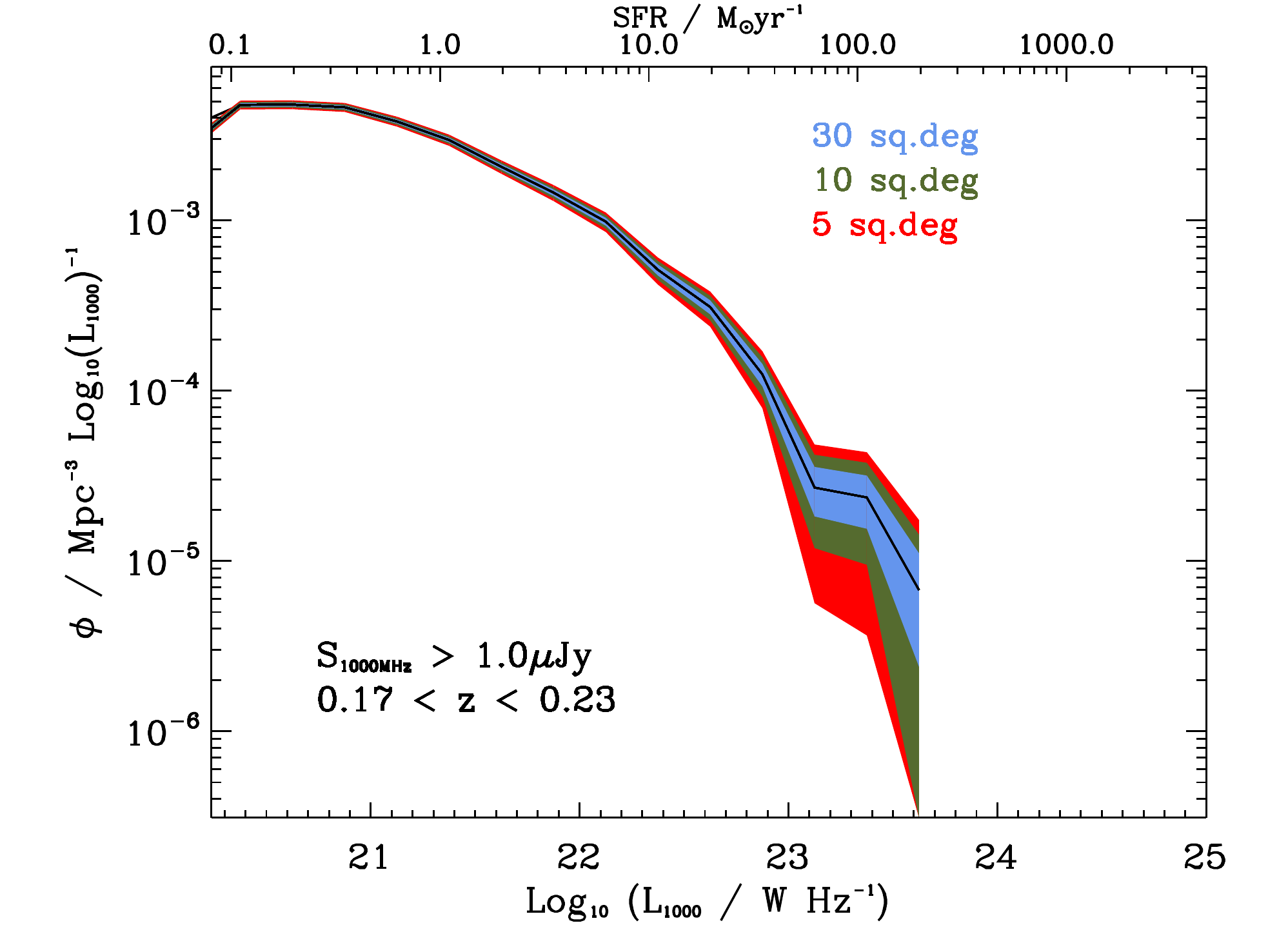}
\includegraphics[width=7.5cm]{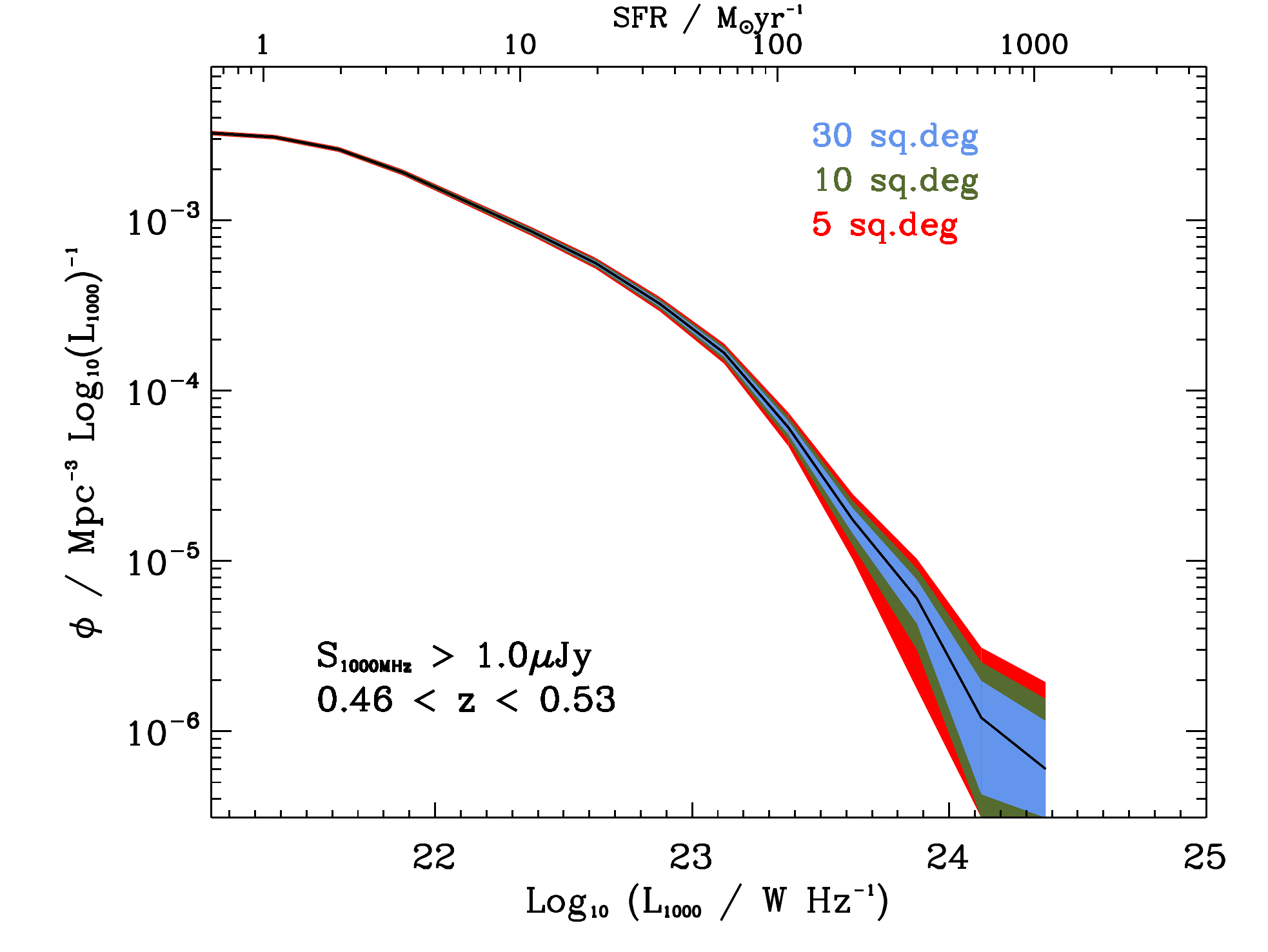}
\includegraphics[width=7.5cm]{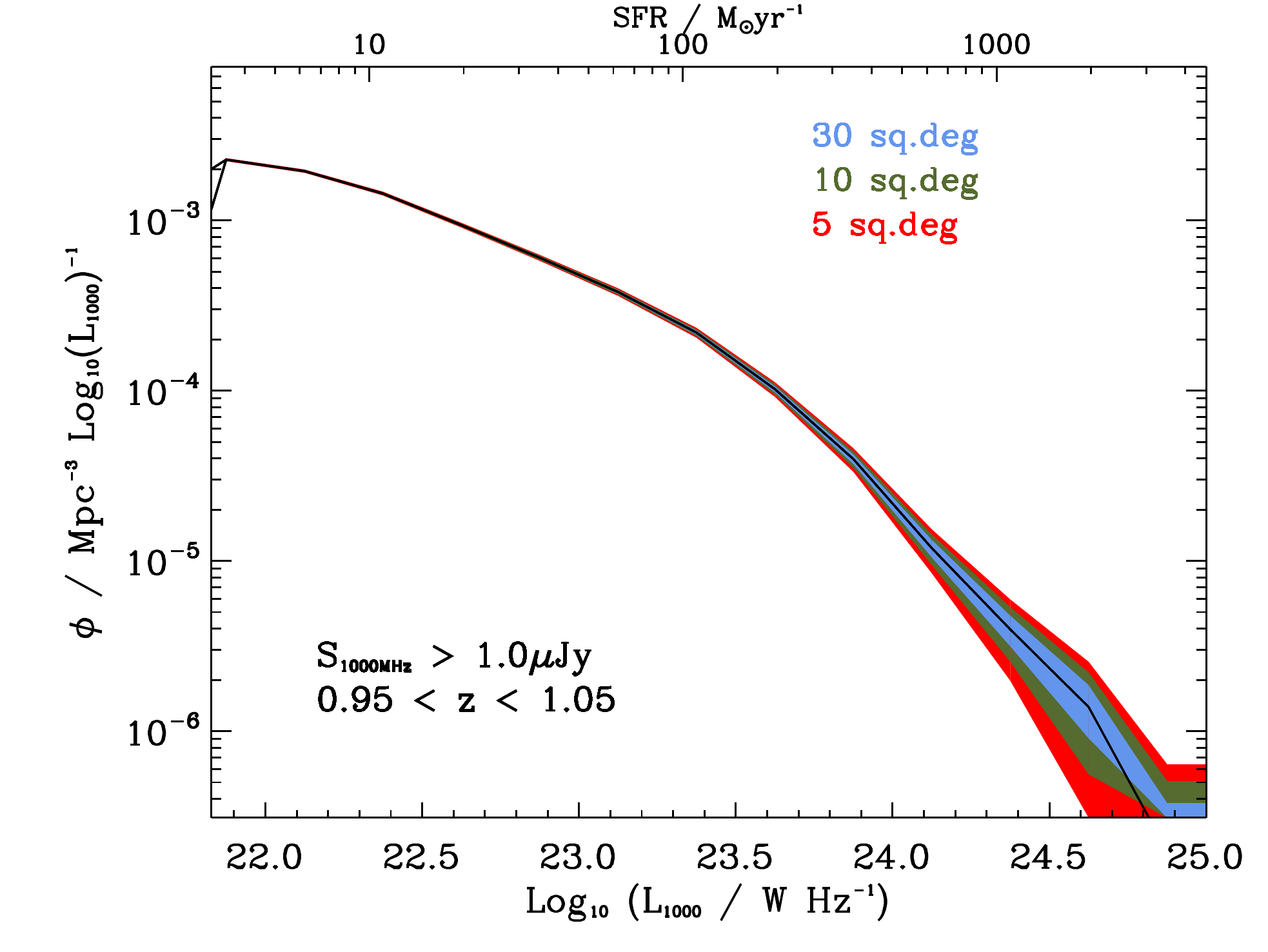}
\includegraphics[width=7.5cm]{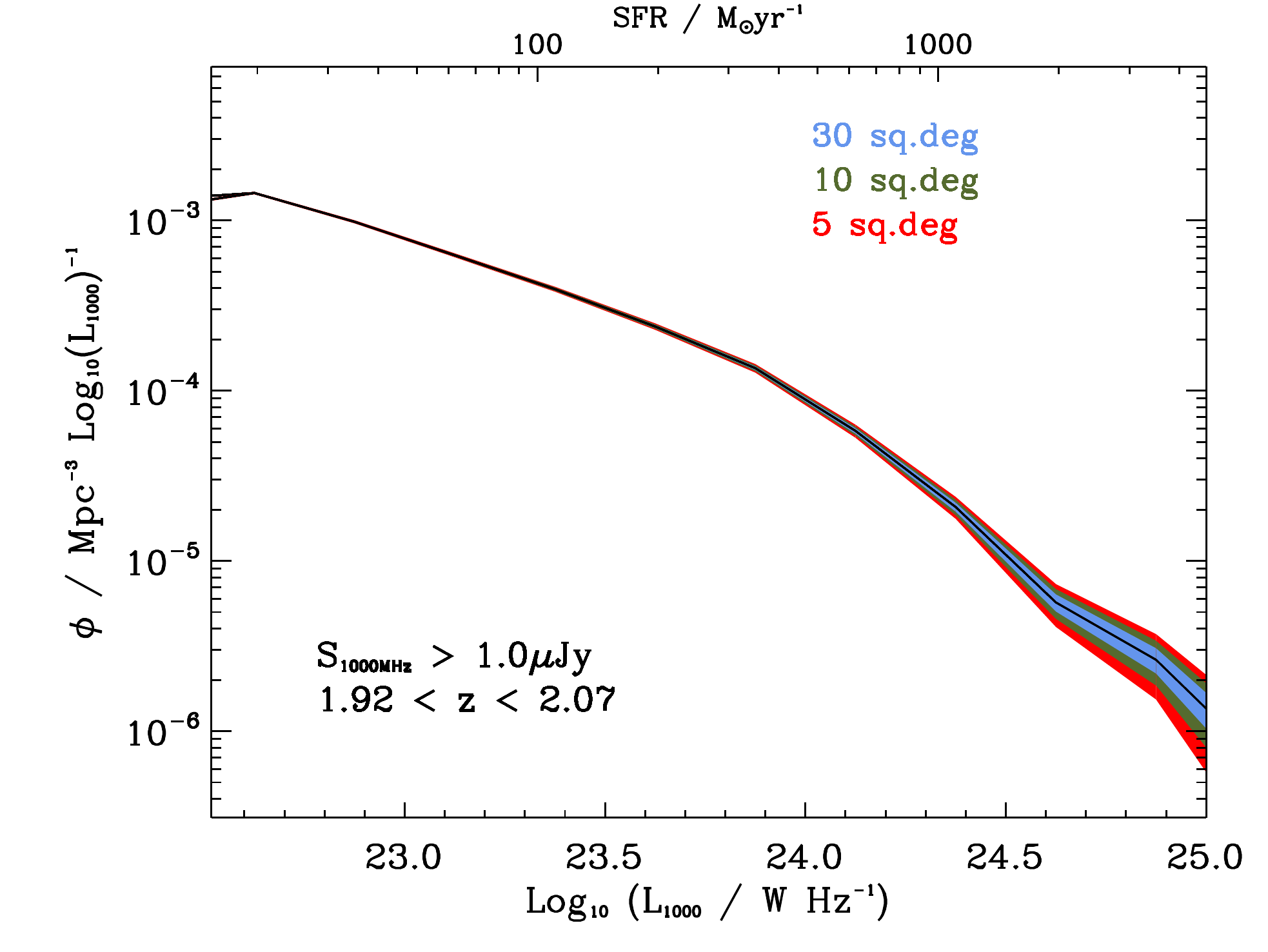}
\includegraphics[width=7.5cm]{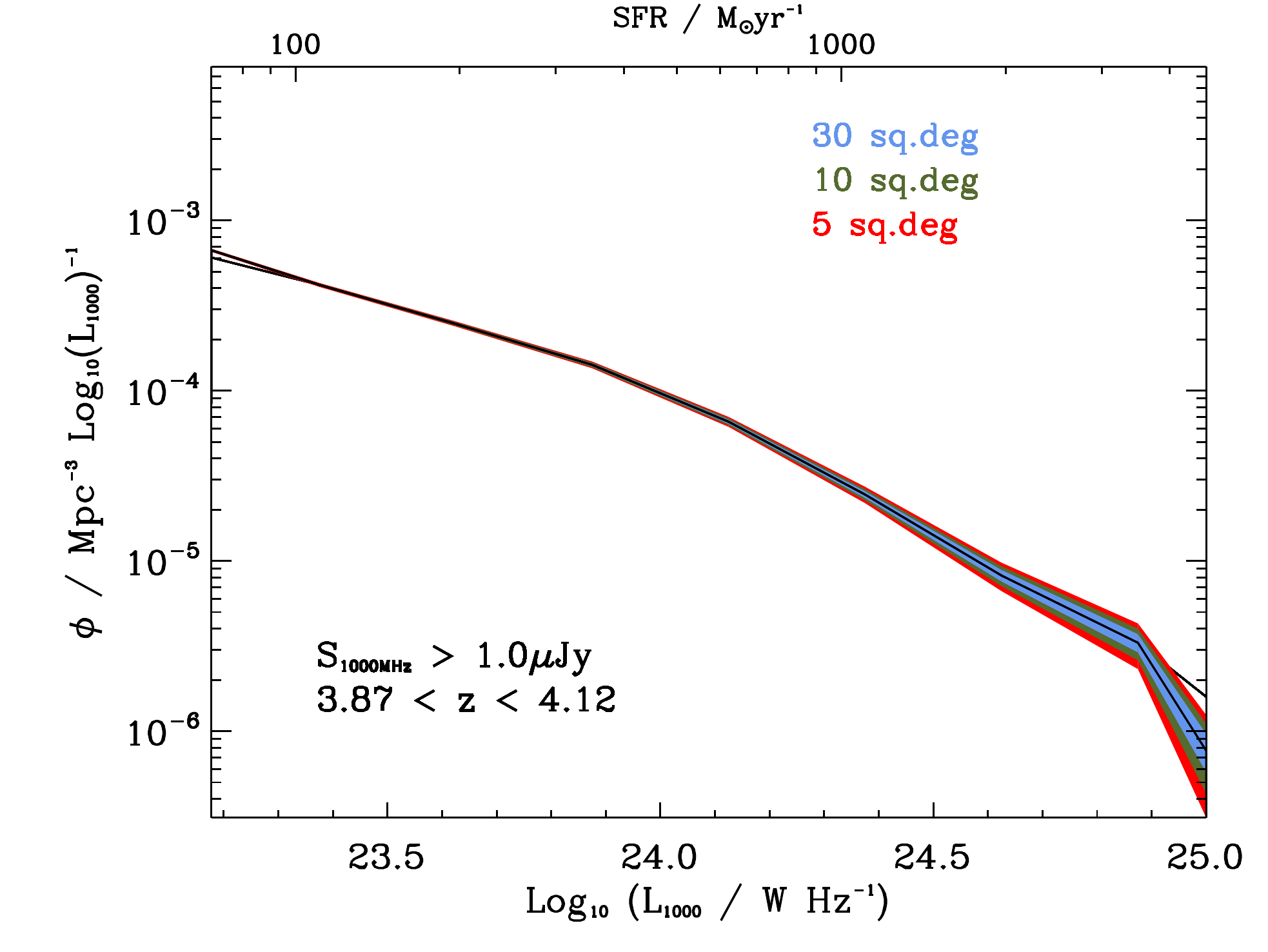}
\includegraphics[width=7.5cm]{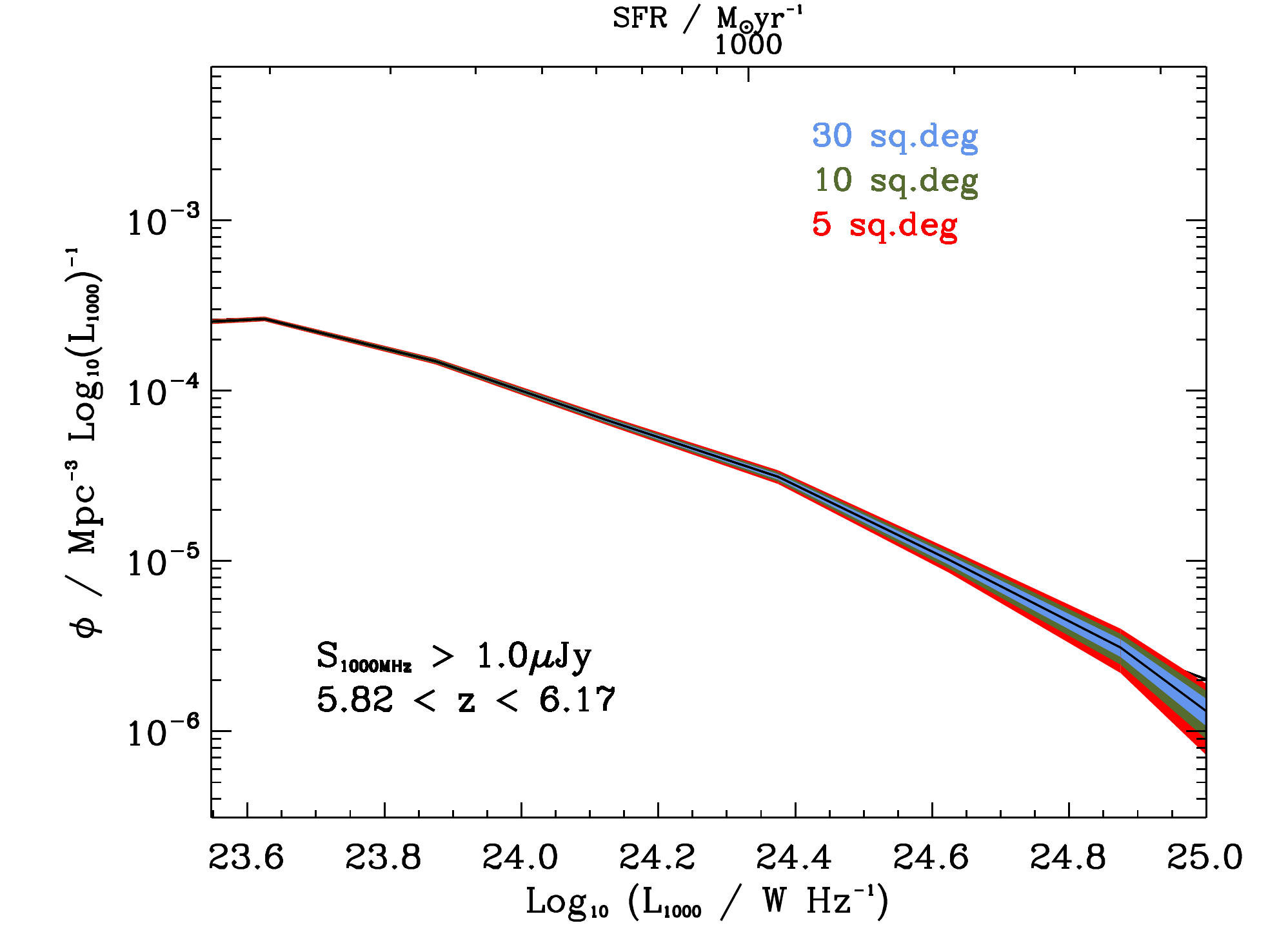}
\caption{Radio luminosity functions (LFs) for star-forming galaxies with $\Delta(\log_{10}L_{1000}) = 0.25$,
  in six redshift bins of width typical of expected photometric
  redshift uncertainties, to a flux-density (5$\sigma$) limit of 1~$\mu$Jy. The red region corresponds to the Poisson
  uncertainties for a 5~deg$^2$ survey, green is for 10\,deg$^2$ and
  blue is for 30\,deg$^2$. The upper
  axis shows the star-formation rate determined from the radio
  luminosity, extrapolated from 1.4~GHz using a spectral index of
  $\alpha = 0.7$.  Note that the range on the abscissa-axis change
  from panel to panel to aid the reader in assessing the uncertainty boundaries. \label{fig:lf_deep} }
\end{figure}

In the following sections and in Figures~\ref{fig:lf_udeep}, \ref{fig:lf_deep} and \ref{fig:lf_wide},
we show the predicted measured radio luminosity function of star-forming
galaxies for three surveys. These are based on the simulations of \cite{Wilman2008, Wilman2010},
assuming a moderate decline in the star-formation rate density at
$z>2$. Together they cover enough cosmic volume from $z =
0\rightarrow 6$ to minimise the Poisson uncertainty, and in the case of
the wide and deep surveys, sample variance (see
Section~\ref{sec:SF-MS}). We note
that the uncertainties on the luminosity function as presented are
entirely Poissonian and therefore depend on the volume
surveyed, which is why the uncertainties in Figure~\ref{fig:lf_udeep}
generally exceed those on the shallower tiers in Figures~\ref{fig:lf_deep} and \ref{fig:lf_wide}.

\subsection{Ultra Deep}

A single deep pointing with the SKA1-MID will be comparable in
size to the deepest fields currently surveyed at other wavelengths,
although the SKA1-MID deep field would be over a somewhat wider area
(1-2~deg$^2$) than the bulk of the ancillary data, which will come
from {\em HST}, {\em JWST} and ALMA, covering optical through to far-infrared wavelengths.
This may also warrant a multi-frequency approach with SKA1-MID, where we
sample from the synchrotron dominant regime at low frequencies through
to the free-free emission that is detectable at higher frequencies, and
where the limited primary beam is not a significant problem
\citep[e.g.][]{Murphy2014}.

The key science for this tier would be to probe the extremely faint
star-forming populations to the highest redshifts (well into the
Epoch of Reionisation). For example, to detect a galaxy with a SFR=20~\Msolaryr at $z\sim7$ would require a 100\,nJy
detection threshold (or an rms of $\sim 20$~nJy). Such a limit would also
allow detection of the star-formation occurring in dwarf
galaxies (M$< 10^{8}$~\Msolar) to cosmologically significant distances
(e.g. $z\sim 0.3$ for a galaxy forming stars at 0.01~\Msolaryr).

\begin{figure}
\includegraphics[width=7.5cm]{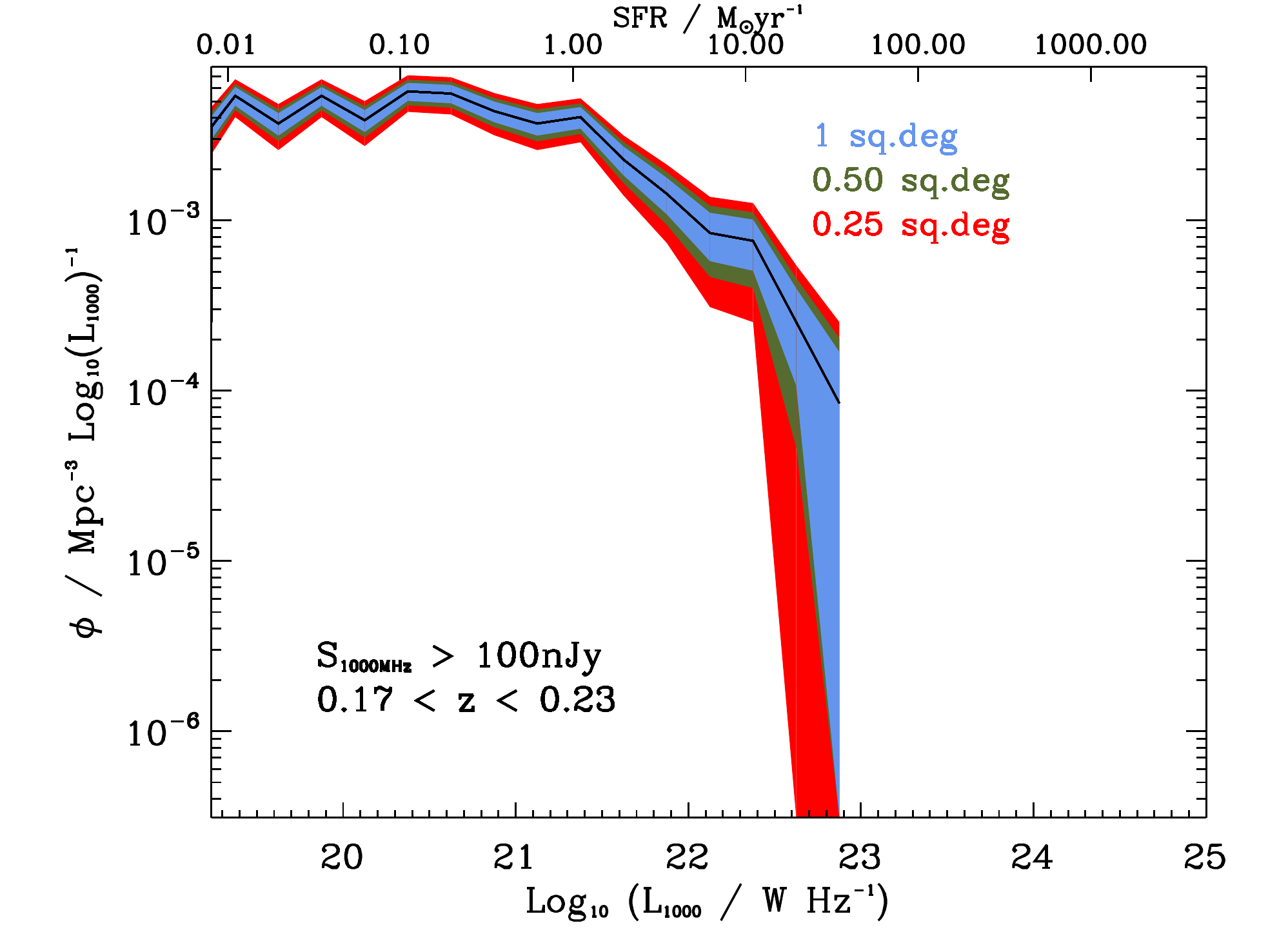}
\includegraphics[width=7.5cm]{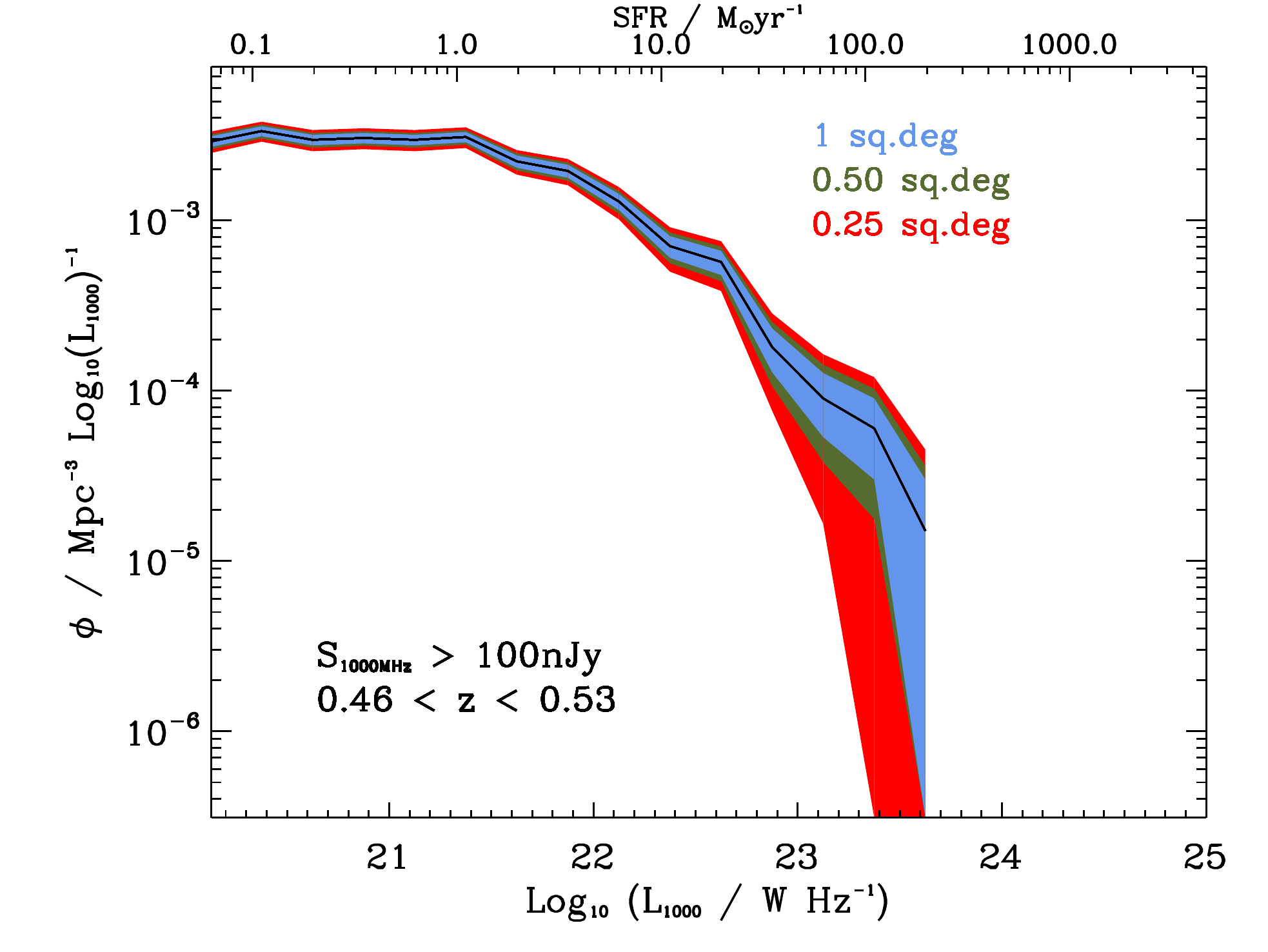}
\includegraphics[width=7.5cm]{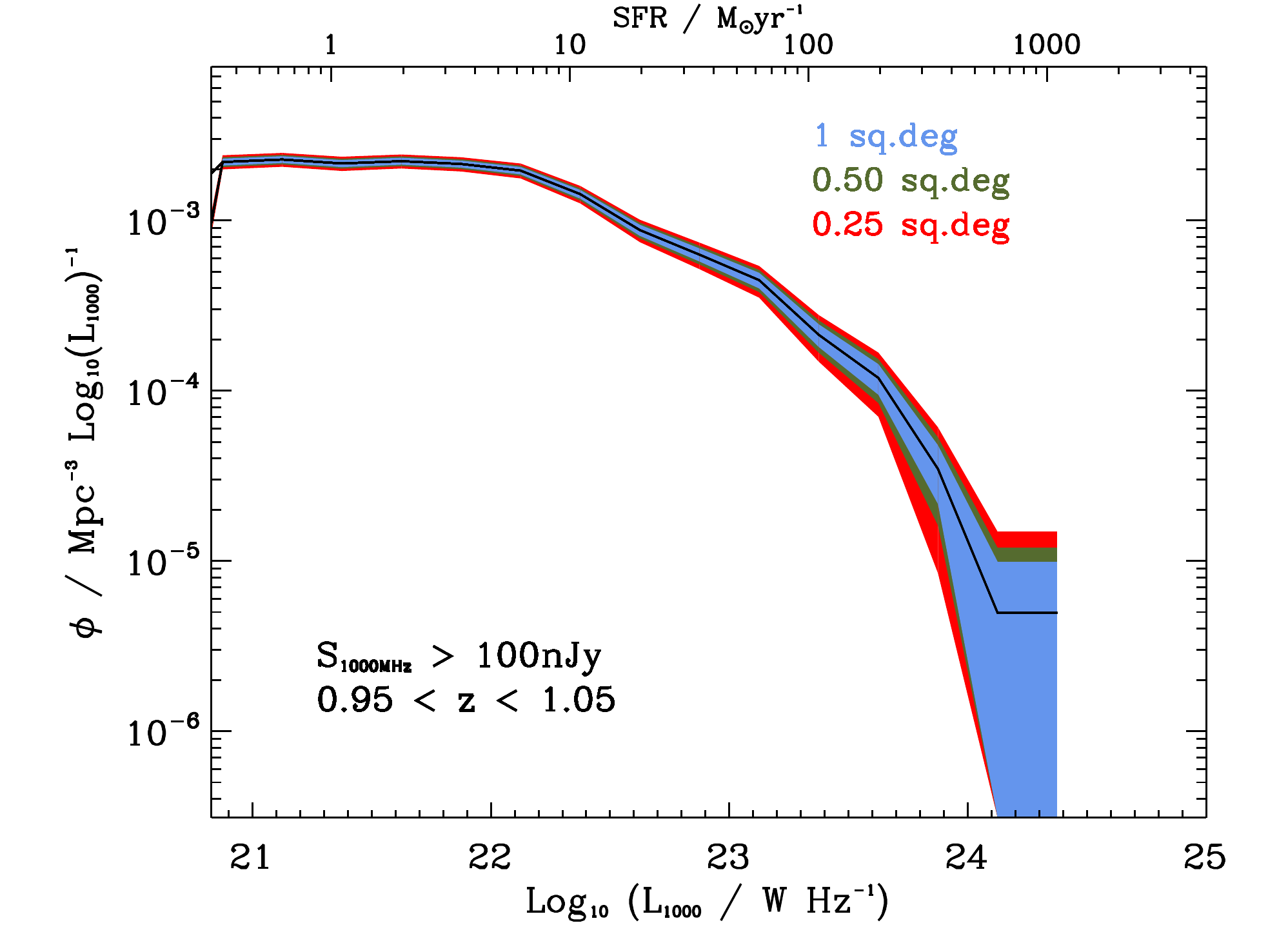}
\includegraphics[width=7.5cm]{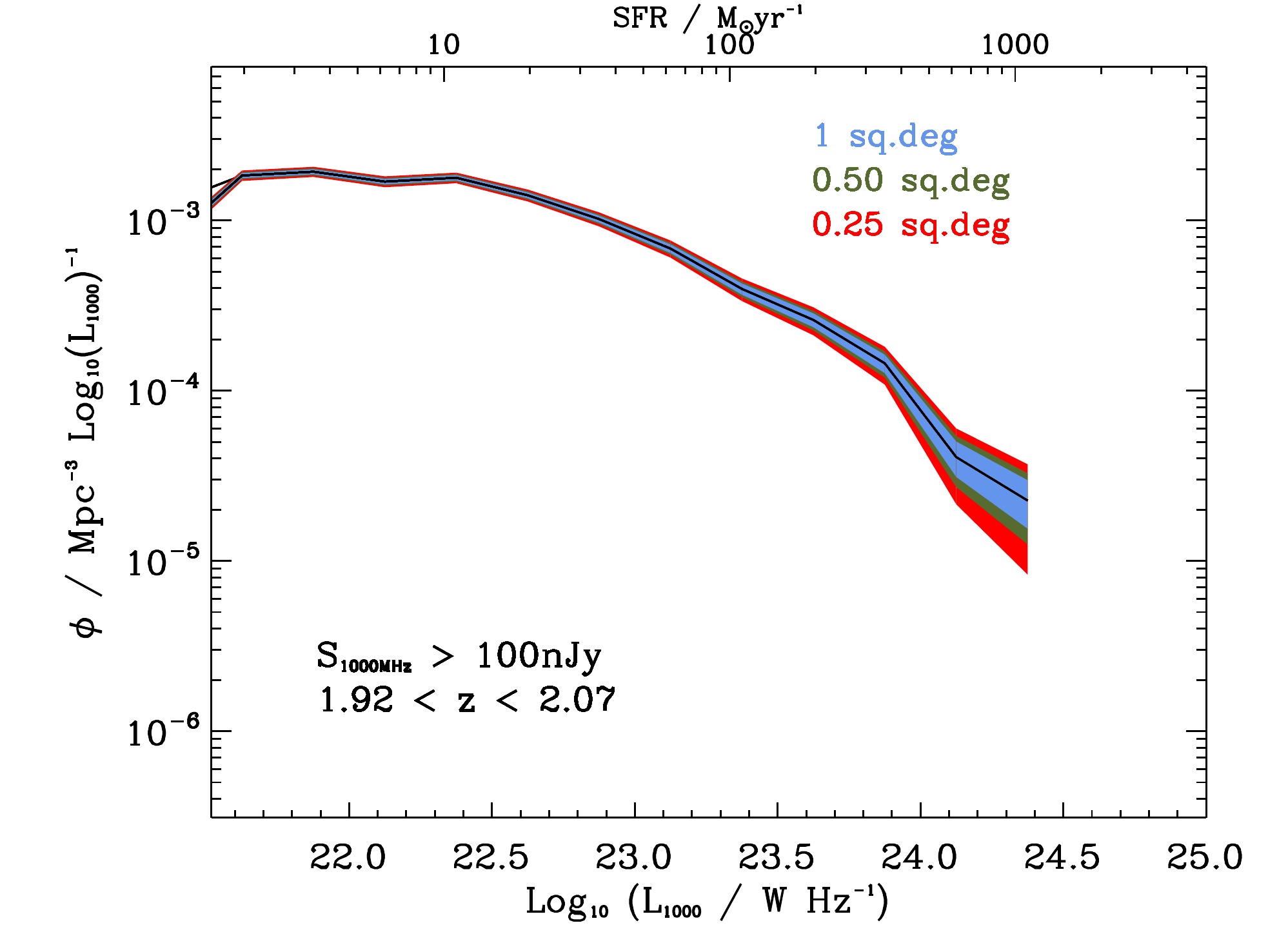}
\includegraphics[width=7.5cm]{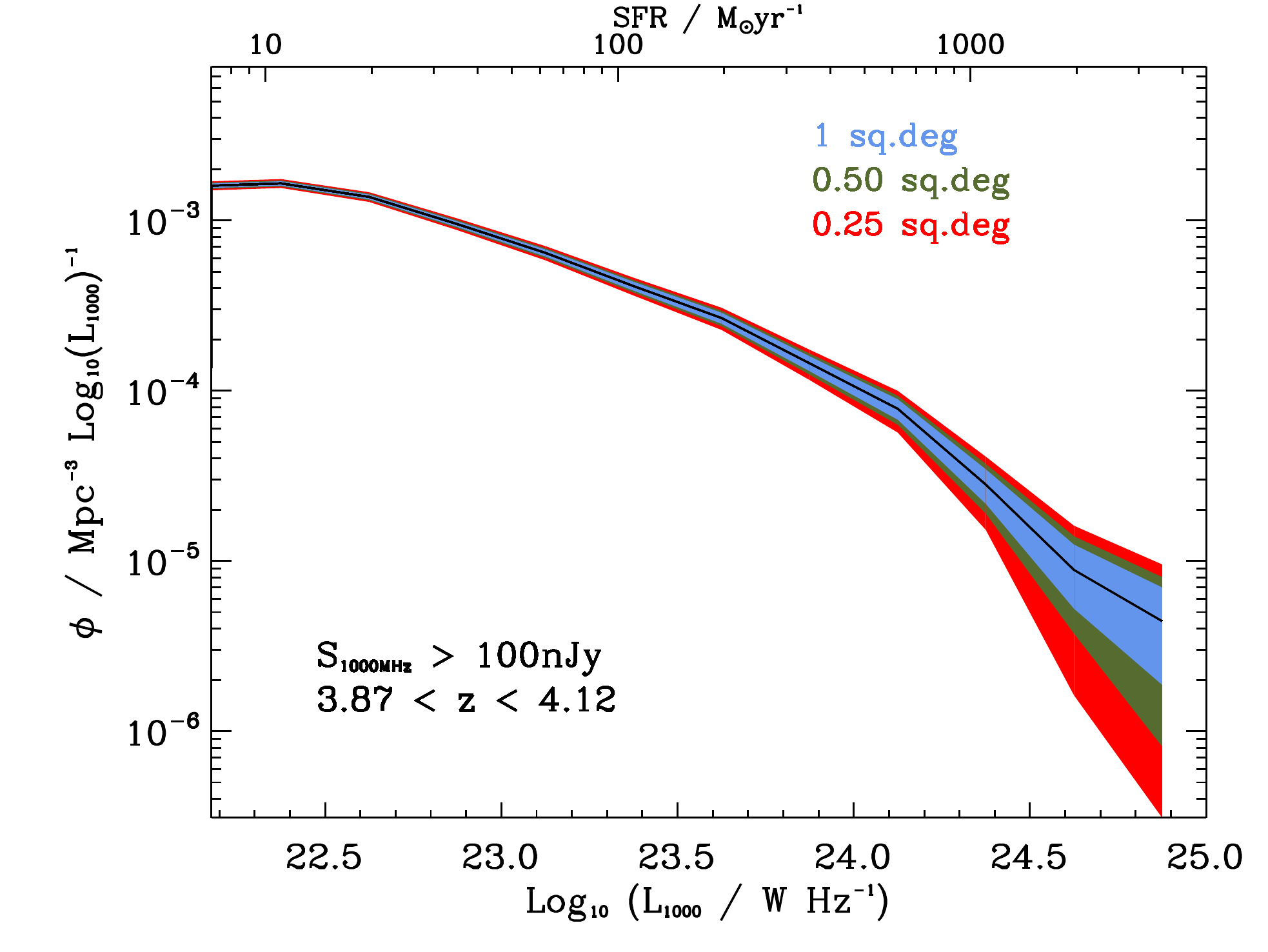}
\includegraphics[width=7.5cm]{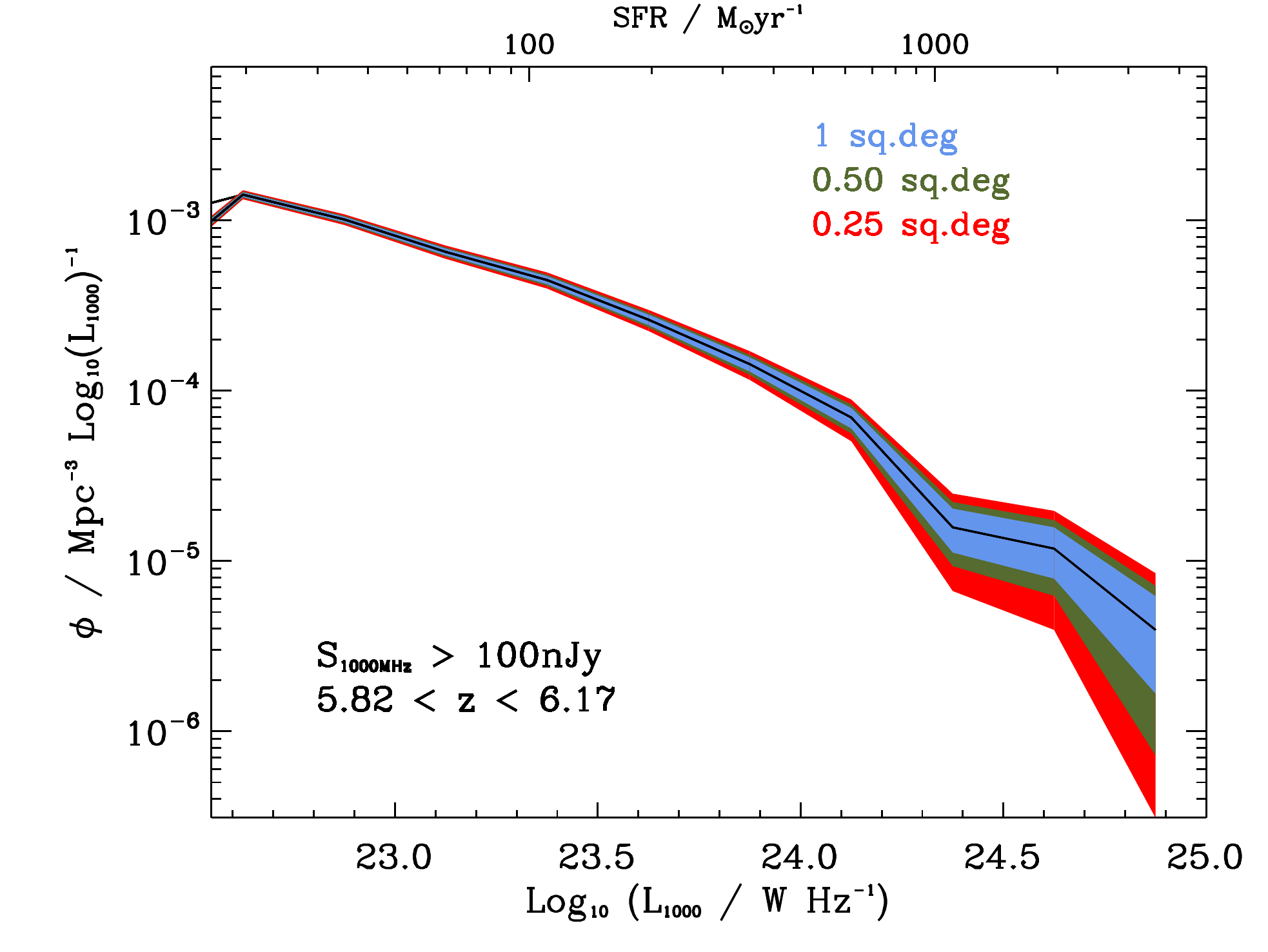}
\caption{Radio luminosity functions (LFs) for star-forming galaxies
  with $\Delta(\log_{10}L_{1000}) = 0.25$,
  in six redshift bins of width typical of expected photometric
  redshift uncertainties, to a flux-density (5$\sigma$) limit of 100\,nJy. The red region corresponds to the Poisson
  uncertainties for a 0.25~deg$^2$ survey, green is for 0.5~deg$^2$ and
  blue is for 1~deg$^2$. The upper
  axis shows the star-formation rate determined from the radio
  luminosity, extrapolated from 1.4~GHz using a spectral index of
  $\alpha = 0.7$.  Note that the range on the abscissa-axis change
  from panel to panel to aid the reader in assessing the uncertainty boundaries. \label{fig:lf_udeep}
  }
\end{figure}

Figure~\ref{fig:lf_udeep} shows the constraints that would be achieved
for
the radio luminosity function from such a survey for three survey
areas. We note that the primary beam of SKA1-MID at 1000~MHz
and 700~MHz are around
0.4~deg$^2$ and 1.5~deg$^2$ respectively, but utilising the full
bandwidth increases the effective sensitivity substantially, at the
cost of a reduced field-of-view at the top end of the frequency band. Therefore, for the
ultra-deep tier we only consider a very small ``single-pointing''
strategy, which means that the central, highest-sensitivity
part of the primary beam can be considered separately to a strategy
that utilises the full area of the primary beam, which naturally has a fall
off in sensitivity aligned with the beam shape. In practice alternative strategies,
which involve some level of mosaicking, should be considered to ensure
a more uniform sensitivity across the preferred survey area.

Figure~\ref{fig:lf_udeep} shows that sample-variance limited constraints
can be made on the evolution of the star-formation in galaxies with SFR$\sim 10$~\Msolaryr~out to $z\sim 4$, and that we can determine the
evolution of galaxies with SFR$\sim 100$~\Msolaryr to $z \sim 8$. Furthermore,
such surveys are feasible over $\sim 1$~deg$^2$, and as such provide an
interesting complement to surveys that will be carried out with the
{\em JWST}, which will have the sensitivity to detect similar galaxies at
near-infrared wavelengths. 

\subsection{Deep}

The role of the deep survey is to provide a census of the Universe
since the epoch of reionisation ($z<6$) through to $z\sim 1$. In order
to probe all environments at these redshifts a survey area of 15-30~deg$^2$ is required, and there are trade-offs in depth versus area that
can be made within this specification. However, given that the key
multi-wavelength data will come from LSST \citep[see
e.g.][]{Bacon2014} and complementary
near-infrared surveys, then it would be sensible to survey the LSST
deep drilling fields, four of which are likely to be the
COSMOS/UltraVISTA, XMM-LSS, CDFS and ELAIS-S1 fields. LSST will provide around 35~deg$^2$ over these fields with contiguous coverage over 9~deg$^2$ patches of sky in each.

A detection threshold sensitive to $\sim 50-100$~\Msolaryr at $z\sim6$,
suggests a flux-density limit of 0.2$\mu$Jy rms over this area. In
Figure~\ref{fig:lf_deep} it is clear that a 30~deg$^2$ survey to a
5$\sigma$ flux-density limit 1$\mu$Jy will provide sample-variance
limited constraints on the evolution of Milky Way-type galaxies to
$z\sim 2$, whilst providing sufficient area to detect the rarest and
more luminous starbursts out to the epoch of reionisation. 
There is evidence for the most massive galaxies at high redshifts to
be more dusty than their lower-mass counterparts
\citep[e.g.][]{Willott2013}. This may give the SKA a unique niche in the study of the
high-redshift Universe, because although such dusty objects could be
detected by far-infrared/submm observatories, the resolution of such
facilities is generally prohibitive to identifying their
optical/near-infrared counterparts. Indeed, radio observations have
been used to associate such sources in the past \citep[e.g.][]{Ivison2007,Heywood2013}. Furthermore, ALMA may be very efficient at studying such
galaxies in detail once they are found, but the small field-of-view of
ALMA essentially precludes it from discovering the rarest and most
extreme galaxies in the early Universe.

\subsection{Wide}

\begin{figure}
\includegraphics[width=7.5cm]{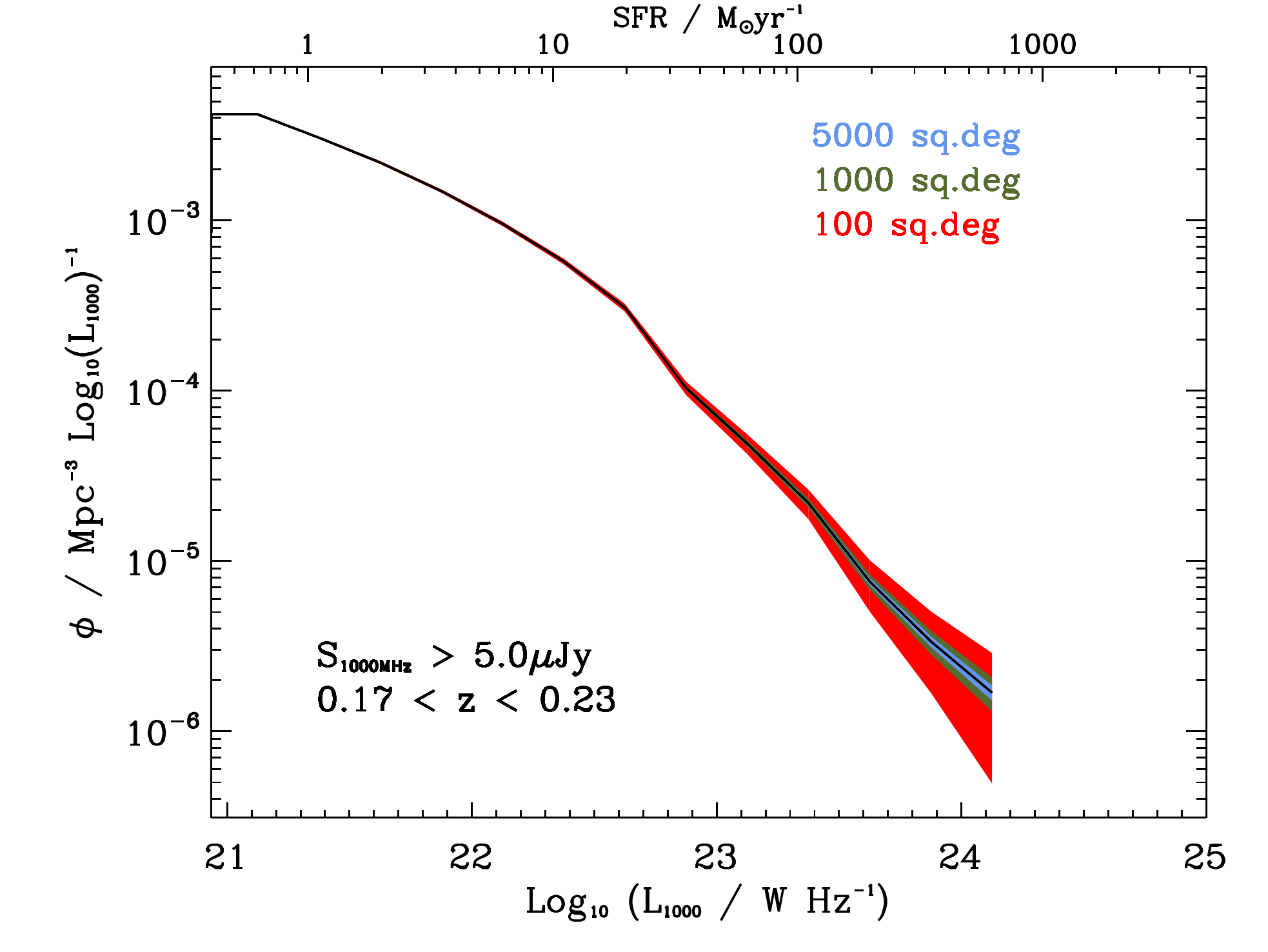}
\includegraphics[width=7.5cm]{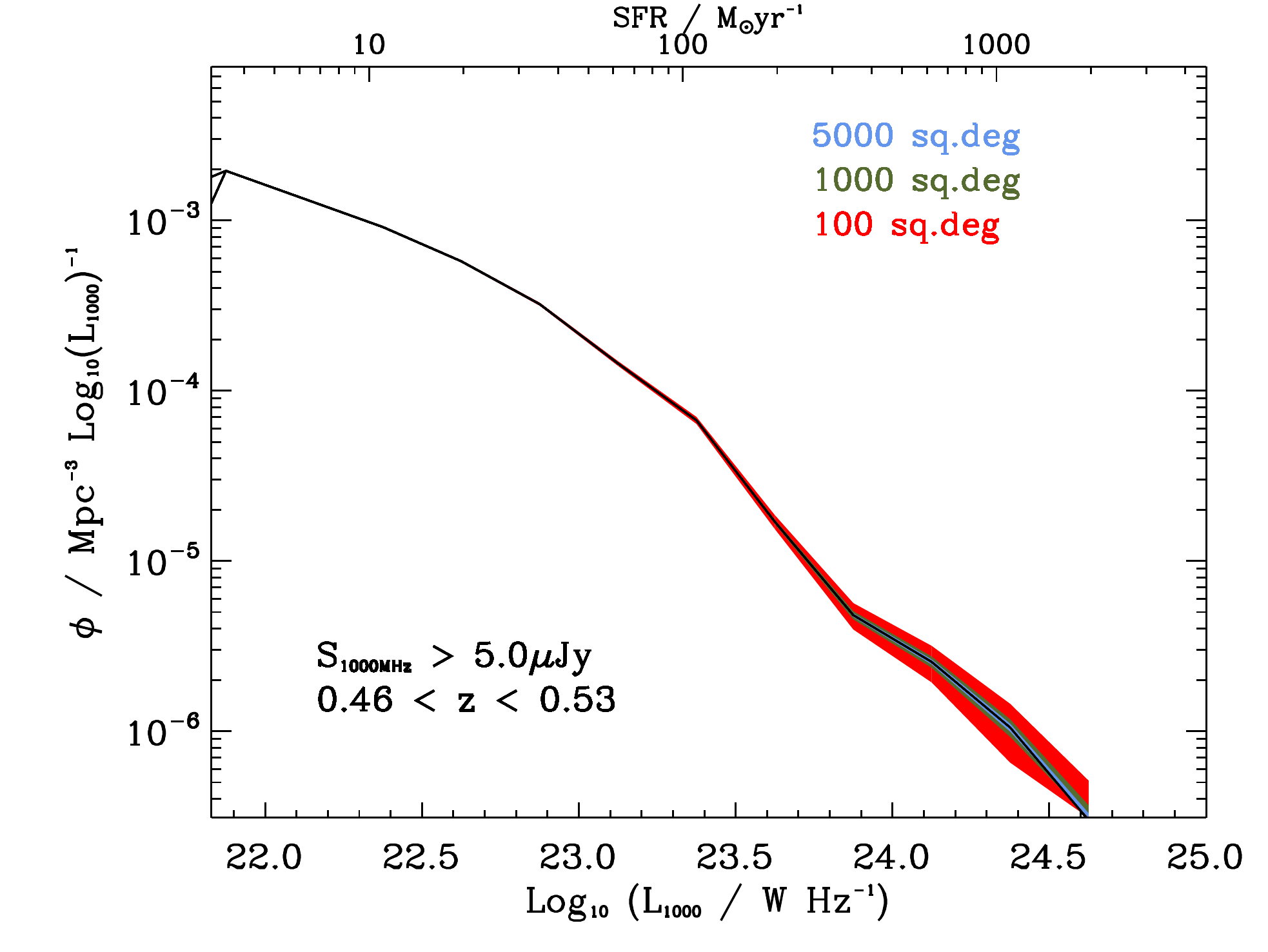}
\includegraphics[width=7.5cm]{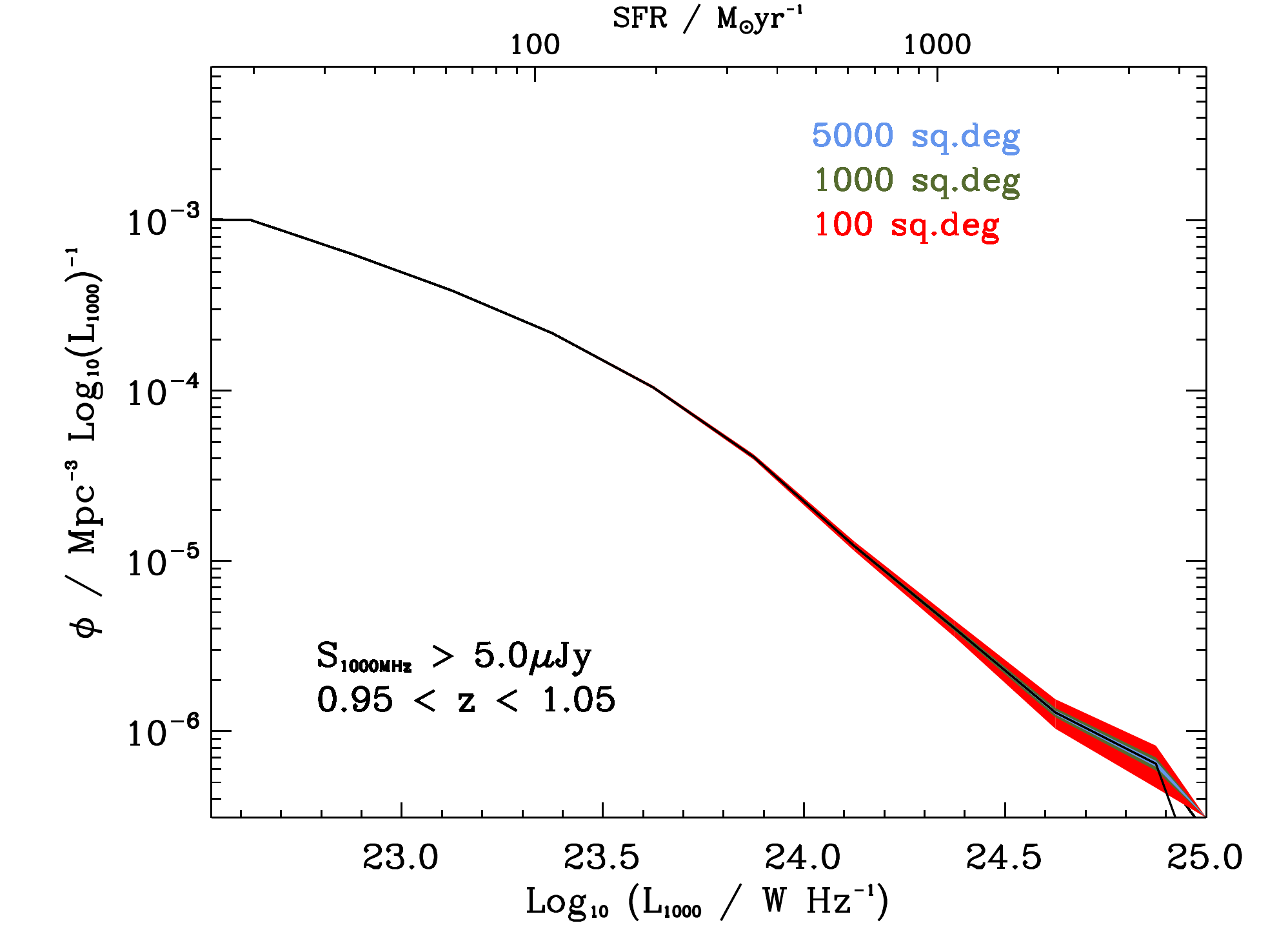}
\includegraphics[width=7.5cm]{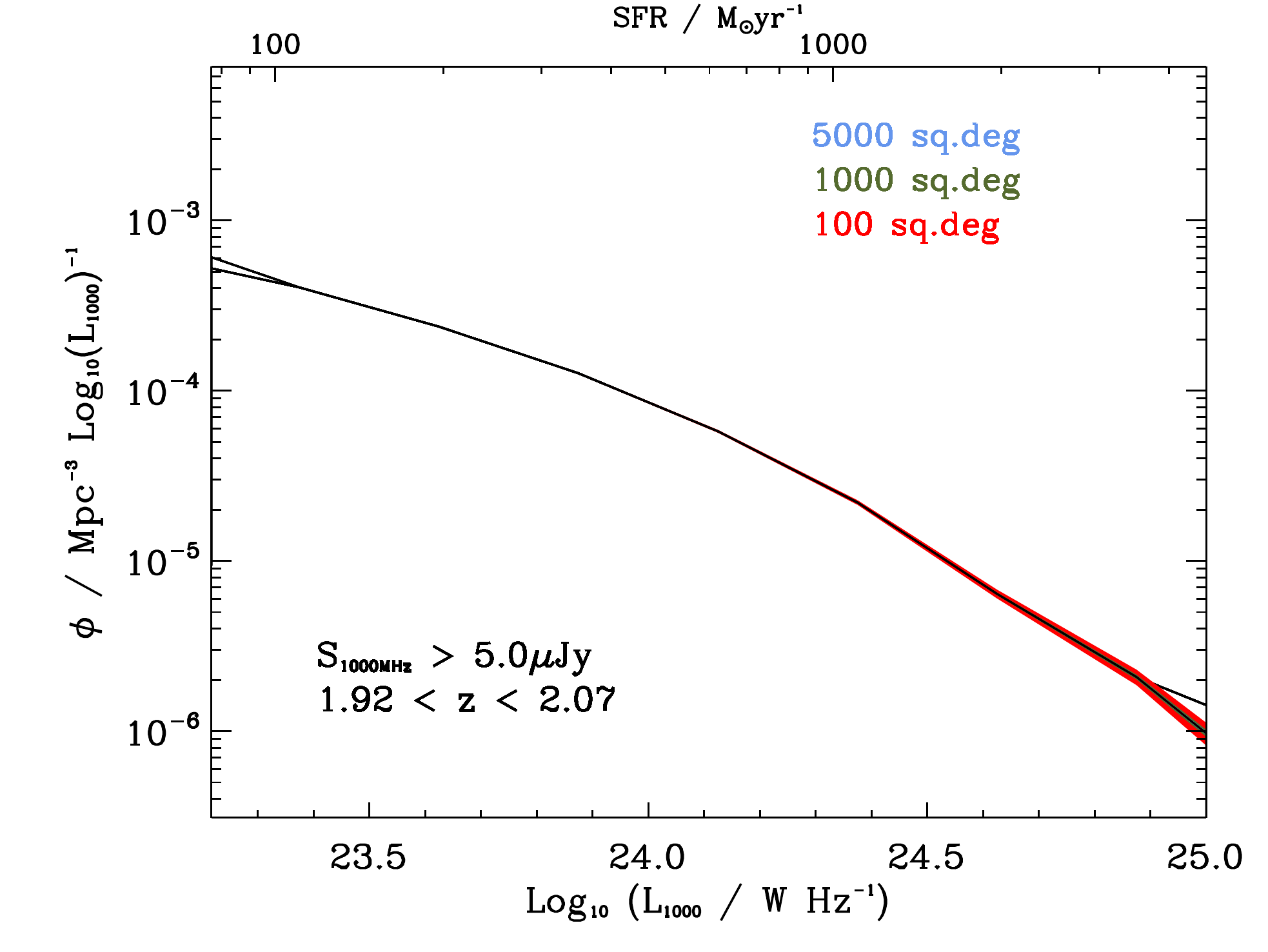}
\caption{Radio luminosity functions (LFs) for star-forming galaxies with $\Delta(\log_{10}L_{1000}) = 0.25$,
  in four redshift bins of width typical of expected photometric
  redshift uncertainties, to a flux-density (5$\sigma$) limit of 5~$\mu$Jy. The red region corresponds to the Poisson
  uncertainties for a 100\,deg$^2$ survey, green is for 1000\,deg$^2$ and
  blue is for 5000\,deg$^2$. Note that the blue and green are invisible in some
  figures due to the Poisson uncertainties being so small. The upper
  axis shows the star-formation rate determined from the radio
  luminosity, extrapolated from 1.4~GHz using a spectral index of
  $\alpha = 0.7$.  Note that the range on the abscissa-axis change
  from panel to panel to aid the reader in assessing the uncertainty boundaries. \label{fig:lf_wide} }
\end{figure}

We also need to relate the findings on the high-redshift Universe from
the deep tier to the lower redshift Universe, therefore we also
support a wider,
shallower tier that will provide a census of the
$z<1$ Universe. To sample the full range of environments at $0.3<z<1$
requires a few thousand square degrees. We again use a
20\,\Msolaryr galaxy at $z=1$ to determine the depth
required. This dictates a depth of around 1$\mu$Jy rms (see
Fig.~\ref{fig:lf_wide}). Based on the existence of ancillary data over
the KIDS/VIKING area (1500~deg$^2$) and the Dark Energy Survey
(5000~deg$^2$) we suggest a combination of these will provide the
necessary ancillary data for the science in this tier, at least until
LSST and {\em Euclid} are well underway.

\section{The star-formation main sequence and the build-up of
  galaxies}\label{sec:SF-MS}

Up until now we have only considered how the evolution of star
formation in the Universe evolves in a general sense, and have not
considered how this may be linked to the evolutionary state of the
galaxy, i.e. how much stellar mass is already in place, and also how
star formation may be related to environmental effects.

The past decade has seen a marked increase in the study of the
relation between the stellar mass and star-formation rate in galaxies,
or when considered together, the specific star-formation rate
\citep[e.g.][]{Erb2006, Daddi2007, Noeske2007}.

The power of the {\em Herschel Space Observatory} has also opened up a new window
on the star-formation history of the Universe, providing us with a
census of obscured star formation from the low-redshift Universe
through to $z>2$. One of the key results to come out of these surveys is
a reinforcement of the relation between star-formation and stellar
mass, the so-called star-formation main sequence \citep[e.g.][]{Elbaz2011,
Magnelli2014, Rodighiero2014}. We can use this link to estimate
how radio continuum surveys with the SKA will be able to provide an
in-depth understanding of the link between the stellar mass build-up,
and the current star formation rate. We assume that the measurement of
the stellar mass in galaxies will come from a combination of optical
(e.g. DES, KIDS, LSST) and 
near-infrared surveys (e.g. VIKING, VIDEO, UltraVISTA) and also  {\em
  Euclid}) \citep[e.g.][]{CiliegiBardelli2014},
i.e. the same data that are also used to determine the photometric redshifts.

\begin{figure}
\includegraphics[width=15cm]{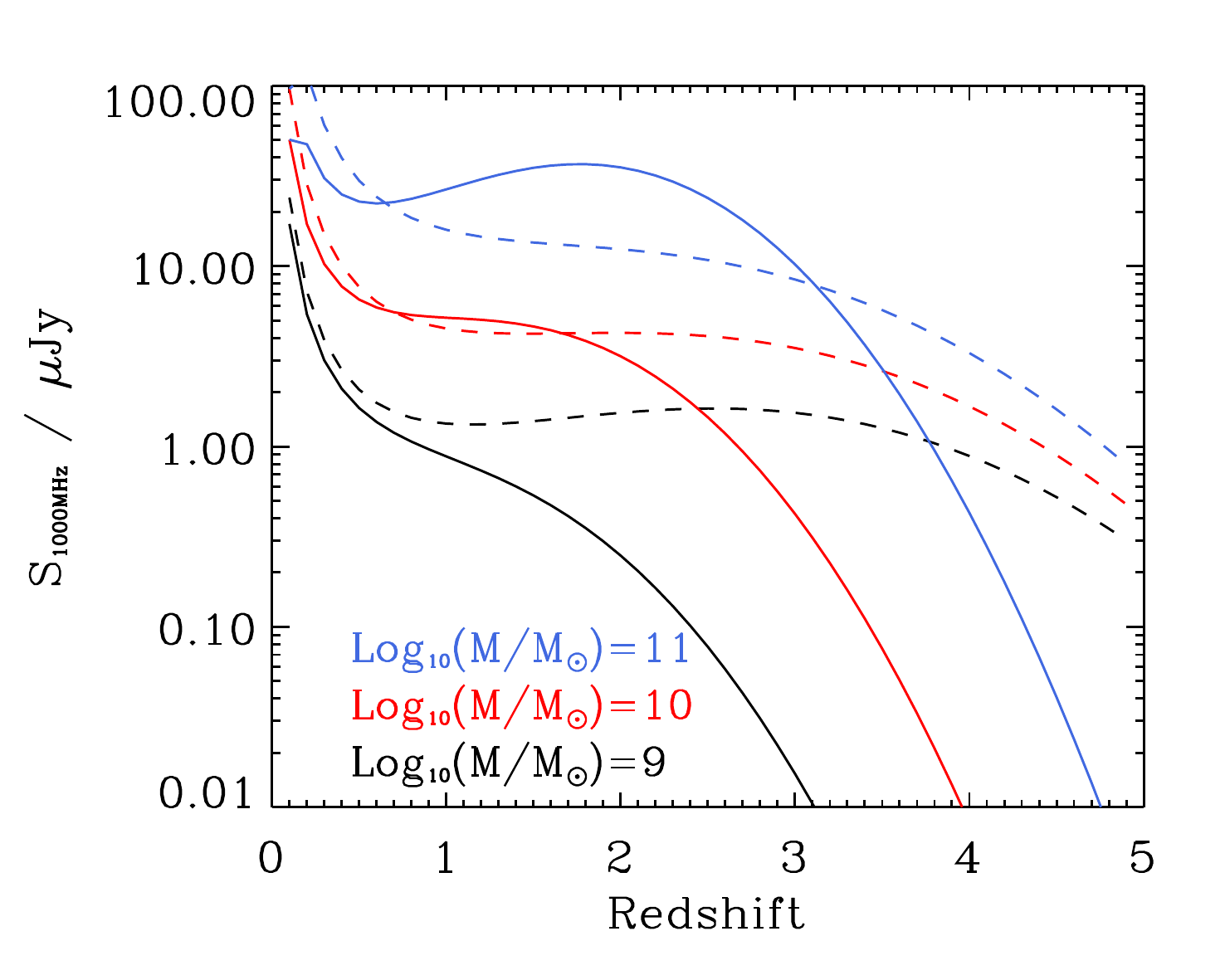}
\caption{The radio flux-density for galaxies of $10^9$, $10^{10}$ and $10^{11}$~\Msolar, that lie on the
  star-formation main sequence as modelled by \cite{Whitaker2012} ({\it
    dashed lines}) and Johnston et al. (in prep.) ({\it solid lines}). Note that
  the form of the relation at $z>2$ is highly uncertain, which can
  explain the large dichotomy between the two prescriptions. \label{fig:SF-MS} }
\end{figure}

In Figure~\ref{fig:SF-MS} we show the predicted average radio flux-density
that would be detected from galaxies of a given mass as a function
of redshift, if they lie on the star-formation main sequence, based on the work of \cite{Whitaker2012} and Johnston
et al. (in prep.). This assumes an intrinsic relationship between the
stellar mass of a galaxy and its star-formation rate, which evolves
strongly with redshift. However, we note that the constraints beyond
$z\sim 2$ are very poor and thus the form of the curves at high
redshift should be considered highly uncertain, and this is borne out
by the divergence in the curves at $z>2$ using the two different
studies. We note that many of the most massive galaxies do not
exhibit such levels of star formation. On the other hand there is also a
significant fraction of starburst galaxies that have higher
star-formation rates than those galaxies on the star-formation main sequence.

Therefore, up to the redshift where the current
data are constrained ($z\sim2$),  a survey flux density threshold of
$S_{1\rm GHz} > 1$~$\mu$Jy is sufficient to determine the SFR of a
$10^9$\,\Msolar~galaxy, if it lies on the star-formation main sequence. Given that such data is free from dust extinction
(cf. ultra-violet/optical measurements) and does not suffer from
confusion (cf. submm/far-infrared imaging), the SKA will provide the
ideal way to push the study of the relation between stellar mass and
star-formation to the highest redshifts.

\begin{figure}
\includegraphics[width=7cm]{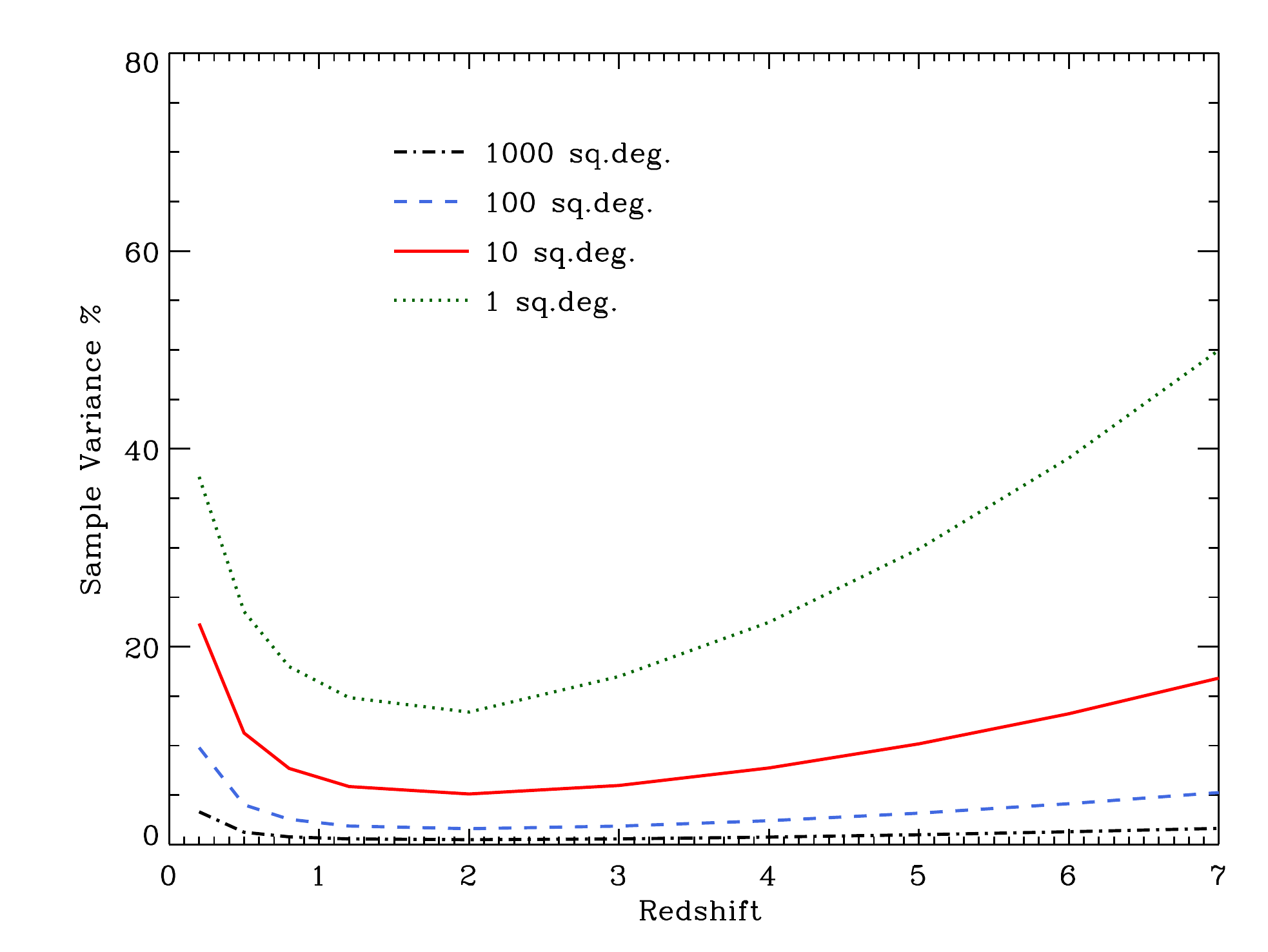}
\includegraphics[width=7cm]{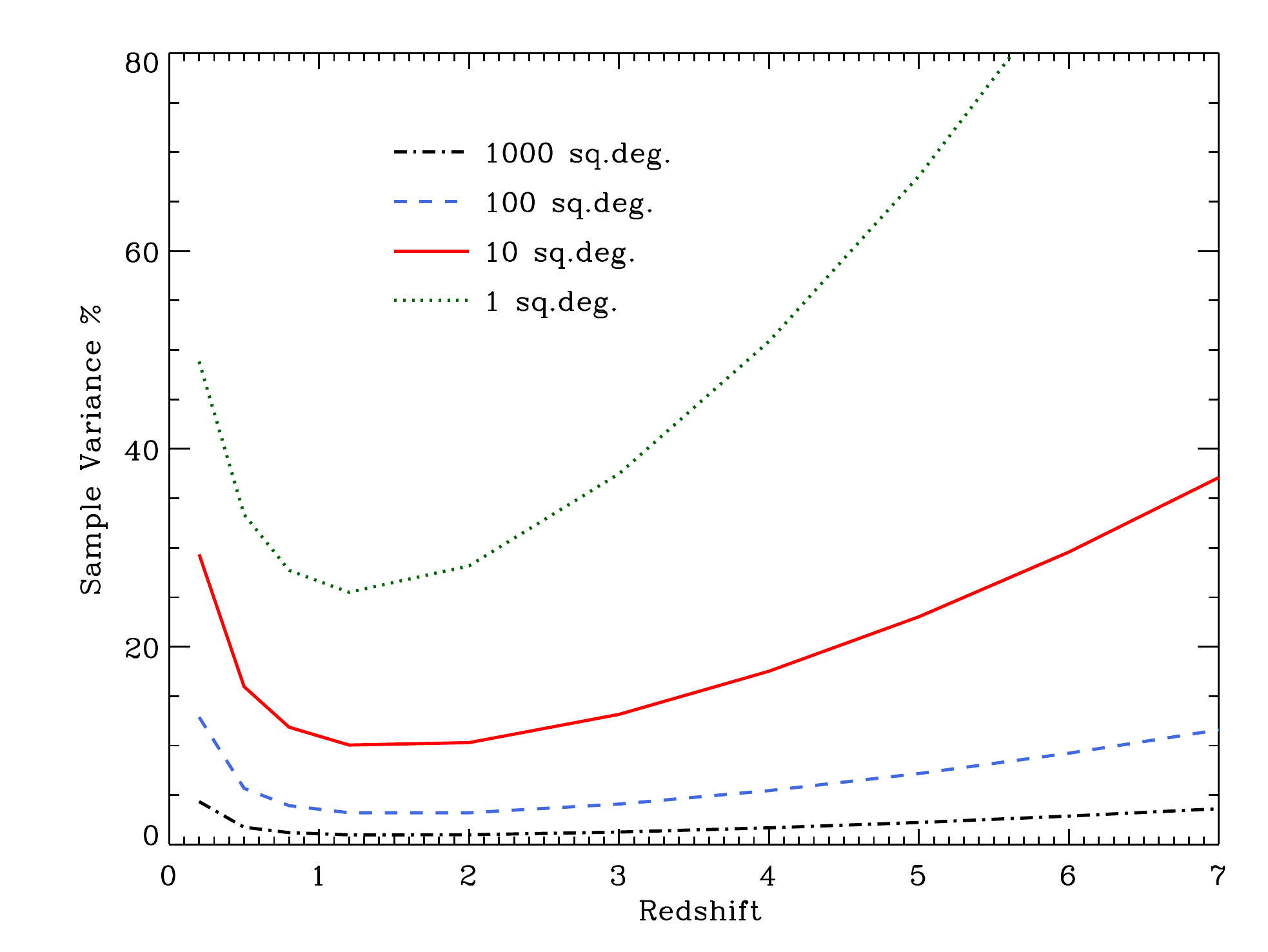}
\caption{Percentage sample variance for massive galaxies; ({\it left})
  $10^{10}$\,\Msolar~and ({\it right}) $10^{11}$\,\Msolar, for four
  surveys covering 1, 10, 100 and 1000\,deg$^2$. Based on the prescription of \cite{Moster2011}.} \label{fig:cosmic_variance} 
\end{figure}

However, when considering the relation between star-formation rate and
stellar mass, one needs to consider whether the survey covers enough
cosmic volume to not be severely limited by sample variance, in particular for the most
massive and highly-clustered galaxies. Therefore, in
Figure~\ref{fig:cosmic_variance} we show the expected level of sample
variance for a given survey area as a function of redshift [of bin
width $\Delta z = 0.05(1+z)$], for two
values of stellar mass, using the prescription of \cite{Moster2011}.

It is clear that in order to overcome significant sample variance for
the most massive galaxies at $z>1$ then a minimum of 10\,deg$^2$ is
needed. Such an area means that the sample variance is of the order of
10 per cent, but increases towards higher redshift. Thus, 30\,deg$^2$
is a more appropriate survey area, and having this area comprised of
sub-areas helps further in overcoming sample variance. Thus targeting
3-4 distinct areas that are part of the LSST deep drilling fields, covering a
total area of around 30\,deg$^2$, would provide the ideal balance
between survey execution time and overcoming sample variance.

\section{The role of environment}

The specific star-formation rate  has been shown to correlate with both mass and
environment \citep[e.g.][]{Baldry2006,Scoville2013}. This implies that an environmental mechanism is actively influencing galaxy evolution within the densest environments through the truncation of star formation.
This raises an important consideration with respect to which correlations are actually dependent on environment and which reflect other correlations that are independent of environment.
Recent studies investigating the relationship between mass and
environment have concluded that the quenching effects of environment
on the star-formation rate in galaxies are separable from the quenching processes traced by
stellar mass. \cite{Peng2010} argue that the quenching of star
formation in passive red galaxies is distinct from the general decline
in the global specific star-formation rate of galaxies since $z\sim
2$. They showed that the specific SFRs of star-forming galaxies are, at most,
a weak function of stellar mass and completely independent of
environment. For passive galaxies however, they conclude that
environment has little impact on the evolution of the most massive ($>
10^{10}$~\Msolar) galaxies, where stellar-mass related quenching
dominates through feedback effects. However, for lower-mass passive galaxies ($<10^{10}$~\Msolar), this quenching is the product of environmental interaction processes.

Therefore, surveys with the SKA, which are both sensitive enough to
gain a full census of star formation in the high-redshift Universe,
but also cover enough cosmic volume to sample the full range in
environmental density, will provide the requisite data to significantly enhance
our understanding of the evolution of galaxies. Therefore a tiered
survey that samples enough cosmic volume at each epoch is required to
address these issues.

\section{Spatially resolving star-forming galaxies}\label{sec:resolution}

Studies using MERLIN, \cite{Muxlow2005} have demonstrated that the typical size of
high-redshift star-forming galaxies is around $0.5-1$ arcsec, similar
to what is found at optical wavelengths. Adopting a spatial resolution at radio observations that is poorer than the complementary data at other wavelengths will weaken the impact of SKA radio continuum observations from a purely multi-wavelength perspective. However, there are extremely strong scientific reasons for requiring high resolution.

 High resolution has been key for determining energetics through brightness
 temperature measurements of sources to distinguish between accretion
 and star formation \citep[e.g.][]{Condon1991}, and to directly
 resolve the impact of jets on star formation in the host galaxy \citep[e.g.][]{McAlpine2014,Makhatini2014}. To understand star
 formation we are required to account for the contribution from AGN, and we can only do this with a spatial resolution that allows us to resolve the global star formation activity in the host galaxy.

Furthermore, it is only by resolving such galaxies in the radio that
we will be able to measure their disk-averaged star formation rates
using a wavelength that is both not obscured by dust or can be
confused with the underlying stellar population. When combined with
observations of the molecular gas content of such systems (e.g., CO
with ALMA), we are able to look for differences in gas depletion times
for sources at high redshift, allowing us to investigate whether the
modes of star formation in galaxy disks is actually different at high
redshift relative to the local universe, for statistically significant
samples of sources.  The SKA will make a huge impact in this area with the requisite long baselines.

Finally, high resolution is key to understanding star formation in the
low-redshift Universe. Although not discussed in this chapter, only
with $<0.5$~arcsec angular resolution continuum imaging capability,
the SKA will be able study the individual components of
star-formation and accretion within local ($<100$~Mpc) galaxies. Such
a capability will allow the SKA to be transformational in this area,
providing a complete census of star-formation and low to high
luminosity accretion powered objects, thus allowing the study of
physics of individual object and the characterisation of the role of these processes more widely within the context of galaxy evolution.

We also note that high spatial resolution is key for other
extragalactic science, in galaxy evolution
\citep[e.g.][]{Smolcic2014,McAlpine2014}, strong lensing \citep{McKean2014} and cosmology
\citep{Ferramacho2014,Brown2014,Jarvis2014cos}.

\section{The multi-wavelength requirements}\label{sec:multiwavelength}

The key quantity that is critical for understanding the star-formation
history of the Universe derived from radio continuum observations is
the redshift of the sources. Radio continuum observations generally
provide no indication of the redshift, therefore we require
ancillary data from a range of other wavelengths. This could be
achieved with broad-band photometry and/or more precise spectroscopic
redshifts from future large-format multi-object spectrographs, or
indeed using the SKA itself for measuring the H{\sc i} 21-cm line. 

\subsection{Spectroscopic redshifts}

Even in 2020 it will not be possible to obtain spectroscopic redshifts
for large areas of sky to the faint limits required to gain a
census of star-forming galaxies. However, future spectrographs on 8-m class
telescopes; e.g. Prime Focus Spectrograph (PFS) on Subaru
\citep{TakadaPFS2014}, the Maunakea Spectroscopic
Explorer\footnote{http://mse.cfht.hawaii.edu} (MSE), and the Multi-Object
Optical and Near-infrared Spectrograph (MOONS)
on the VLT \citep{CirasuoloMOONS2012} may provide the survey power to
gain a very good census of the radio sources in the deeper fields. In
particular, at $z<1.2$ and $z>2.2$, PFS and MSE will have the spectral
coverage at visible wavelengths
to obtain emission-line redshifts based on the usual star-formation
tracers, e.g. [O{\sc ii}], H$\alpha$ etc. Both are situated in the
northern hemisphere so will not be able to cover the whole of 
the SKA sky, however in terms of the deep fields suggested, only
ELAIS-S1 at a declination of $<-40$~deg would be difficult to observe.  Moreover, obscured systems that we detect at radio wavelengths but not
at optical wavelengths will still be a problem. The proposed
near-infrared multi-object spectrograph for the VLT (MOONS) could fill in some of this parameter space, with the redshift desert ($1.2<z<2.2$) difficult to access with optical spectrographs.

For wider and shallower surveys, the proposed 4MOST spectrograph,
aiming to survey the entire southern sky in spectroscopy to $r<22$
would provide a basis for obtaining redshifts for the brighter
star-forming galaxies, predominantly in the low-redshift Universe. If
4MOST adopted a tiered survey (e.g. WAVES) whereby the integration time was well-matched to the survey strategy for the SKA continuum survey, then it could fill the gap between the ultra-deep pointings one might expect to carry out with MOONS and PFS, and the wide-area tiers.

The additional benefit of spectroscopy over imaging is that the
emission lines can be used to determine the level of AGN activity
\citep[e.g.][]{JacksonRawlings1997,Herbert2010} or
star formation \citep[e.g.][]{BPTdiagram,Kewley2013} in the galaxy, complementing the radio data.

\subsection{Photometric redshifts}

The majority of the radio sources detected at these faint levels will be
too faint at optical wavelengths to obtain spectroscopic redshifts. We are
therefore reliant on photometric redshifts based on the deep imaging
data that will be available on the same timescale as the SKA. In the
early phases this will be from surveys that are currently underway,
such as COSMOS/UltraVISTA \citep{Scoville2007,McCracken2012}, SXDF/UDS
\citep {SXDFsurvey,Foucaud2007} and the
VIDEO survey fields \citep{Jarvis2013} for the deep surveys, and
KIDS/VIKING \citep{KIDS,VIKING}, DES/VHS
\citep[e.g.][]{Banerji2014} and {\em WISE} \citep{WISE} for the wider areas.

As we move to the full operation of SKA1 then we should also have LSST
and {\em Euclid} imaging, which will provide very deep imaging from the
$g$-band through to $H$-band across a large swathe of the southern
sky. The expected photometric redshift accuracy from such surveys is
$\Delta z \sim 0.05(1+z)$ and we have assumed this in our predictions
for the luminosity function evolution based on the SKA continuum
surveys. However, we note that it is impossible to quantify
the accuracy of photometric redshifts of objects that are fainter than
the limits possible with spectroscopy. Furthermore, emission-line galaxies are generally more difficult
to estimate the photometric redshifts for, due to the uncertainty
surrounding the strength of emission lines which pass through various filters.

In addition to the photometric redshifts, these surveys provide
the necessary data from which to derive other properties of the
galaxies, e.g. stellar mass, optical reddening and morphology. 

\section{The SKA}\label{sec:SKA}

In this section we discuss the technical requirements of the science
presented in this chapter.

The key argument revolves around the need to have
high resolution, in order to avoid the confusion limit and allow the
characterisation of the radio sources based on their morphology (Section~\ref{sec:resolution}),
whilst also aiming to observe at relatively low frequency to maximise
the observed flux density of the sources, due to the steep synchrotron-emission
spectrum ($S_{\nu} \propto \nu^{\-0.7}$ for star-forming galaxies). This is particularly
pertinent at high redshift, where the synchrotron spectrum may steepen
towards high-frequency due to synchrotron losses off CMB photons
\citep[e.g.][]{Murphy2009}. The adopted central frequency of
1000\,MHz, used in this chapter corresponds to a resolution of $\sim
0.4$~arcsec for SKA1-MID, whereas SKA1-SUR would
provide a resolution of $\sim 1.5$~arcsec, i.e. much larger
than the typical angular extent of galaxies at $z>0.5$. 

Furthermore,
given a typical spectral index of $\alpha = -0.7$\footnote{$S_{\nu}
  \propto \nu^\alpha$.}, then sources that are detected at 20$\sigma$ at
700~MHz will be detected at $\sim 12\sigma$ at 1.4~GHz in the same
receiver band, with a resolution
of $\sim 0.3$~arcsec with SKA1-MID (compared to 1~arcsec for SKA1-SUR). This will allow detailed morphologies to be measured
for a large fraction of the sources using SKA1-MID, and certainly to a better
accuracy than non-AO assisted ground-based optical imaging, although
obviously the primary beam is smaller at 1.4~GHz compared to 700~MHz
so a different survey strategy would be required to obtain uniform
coverage at the higher frequency (we note that this would apply to
SKA1-SUR as well if the PAF was optimised towards the lower frequency
end of the band using the maximum number of beams).
If SKA1-SUR was to be used then we would lose all of our ability to
obtain morphological measurements of star-forming galaxies at high
redshift, where the typical size is of order $0.5-1$~arcsec, and
therefore much of the unique science that can be achieved with the SKA.

We also note that SKA1-MID is a faster survey instrument than SKA1-SUR at the
required resolution ($\sim 0.5$~arcsec) at all frequencies below
1.4~GHz. For a fixed resolution this advantage increases with
decreasing frequency. As a resolution of 0.5~arcsec is essential for this
science, then the number of sources is maximised by going to the lowest frequency.


Therefore, the combination of longer maximum baselines and higher instantaneous
sensitivity makes SKA1-MID the preferred facility for this
science case, as the higher surveys speeds at
high frequencies for the PAF technology is negated for the majority of
radio continuum science, due to ability to move to lower frequency with
single-pixel feeds. For the shallower tier SKA1-SUR is more competitive, and
the key limiting factor is the angular resolution.

\subsection{Towards SKA1}

Given that the SKA will be built up over the coming decade, in this
section we highlight the preferred build-out strategy for the science
case outlined above. 

In order to make the most informative surveys as the SKA is expanded,
then enhancing the ability to reach the full depth at the full
resolution as quickly as possible will ensure that the new continuum
surveys for studying the star-formation history of the Universe will
surpass what is possible with the JVLA
\citep[e.g.][]{HalesVLASS,BrownVLASS,JarvisVLASS}. Initially this
should be done at the expense of survey area. Given that the requisite
multi-wavelength surveys will not exist to the required depth over the
survey areas described in Section~\ref{sec:rlf}, then it makes perfect
sense to start the surveys on smaller scales, but retaining the final
depth.
As such, for 50 per cent of SKA1-MID, surveying the 13.5~deg$^2$
covered by the combination of the VIDEO and UltraVISTA near-infrared
surveys would provide the greatest leap in our understanding of the
total star-formation rate in galaxies, over the epoch where the
Universe was undergoing its most active phase at $1<z<4$. This would
also enable the study of the environmental dependence of star formation over the vast
majority of environments, and provide sufficient area in four
independent fields to reduce the significant effects of sample variance.

 Given that
low-surface brightness sensitivity is not a key element for such
surveys, then this means that in order to obtain the requisite
resolution, the core of SKA1-MID could be delayed with respect to
the long baselines, which are critical to retain the resolution for
morphological studies and to avoid confusion at these very deep
levels. This would require a similar time to complete as the full
30\,deg$^2$ survey, as the poorer sensitivity is balanced by the reduced
survey area.

\subsection{SKA1}

The three surveys in Section~\ref{sec:rlf} could be carried out fully with
SKA1-MID, providing a unique census of star formation
from the local Universe through to the epoch of reionisation. The key
elements of SKA1-MID are the high sensitivity, large bandwidth in Band
1 or 2, along with the high resolution at relatively low frequency, which
are required to push to low star-formation rates at high redshift
whilst retaining morphological information.

\subsection{SKA2}

SKA2, with a factor of 10 increase in sensitivity and
resolution, will be unrivalled for studying the total star
formation rate in galaxies. Although at this stage it would be
possible to extend the deep field and ultra-deep fields described in
Section~\ref{sec:rlf} to much wider areas, there may be little
gained if this was seen as the default approach. By the mid-2020s our
understanding of the evolution of star formation will have changed
significantly, and it is really the new parameter space that is most
likely to add to our understanding. For example, obtaining {\em JWST}-like
resolution over enough cosmic volume that all environments are probed,
out to the highest redshifts, would be a significant advancement over
the foreseeable surveys with SKA1-MID. Therefore, the highest priority
in our opinion would be to; a) cover the 30\,deg$^2$ where the best
ancillary data lies, to the depth of the ultra-deep survey described
above, but at a resolution of 30\,milli-arcseconds. This would allow
the study of  star forming regions on the scale of 200~pc (i.e. the size of the
Tarantula Nebula in the LMC) up to the highest redshifts, and, b) conduct ultra-deep surveys of dense regions to better understand the role of environment.

\section{Conclusions}

The SKA promises to be the premier facility for understanding the
evolution of star formation in the Universe. Unlike optical and
ultra-violet observations, the radio emission is not extinguished by
dust, and thus provides a unique method to trace the total
star-formation rate in galaxies. Furthermore, the resolution that is
possible with the SKA surpasses what will be possible with the {\em JWST},
allowing morphologies and individual star-forming regions to be
observed to the highest redshifts.  Such observations, which will also
cover enough area of sky to overcome sample variance, and allow
star-formation to be studied as a function of galaxy environment, will
provide the best method for understanding the build up of stellar mass
in the Universe. 

\bibliographystyle{apj_long_etal}
\bibliography{matts2}

\begin{thebibliography}{87}
\expandafter\ifx\csname natexlab\endcsname\relax\def\natexlab#1{#1}\fi

\bibitem[{{Afonso} {et~al.}(2005){Afonso}, {Georgakakis}, {Almeida}, {Hopkins},
  {Cram}, {Mobasher}, \& {Sullivan}}]{Afonso2005}
{Afonso}, J., {Georgakakis}, A., {Almeida}, C., {Hopkins}, A.~M., {Cram},
  L.~E., {Mobasher}, B., \& {Sullivan}, M. 2005, \apj, 624, 135

\bibitem[{{Appleton} {et~al.}(2004){Appleton}, {Fadda}, {Marleau}, {Frayer},
  {Helou}, {Condon}, {Choi}, {Yan}, {Lacy}, {Wilson}, {Armus}, {Chapman},
  {Fang}, {Heinrichson}, {Im}, {Jannuzi}, {Storrie-Lombardi}, {Shupe},
  {Soifer}, {Squires}, \& {Teplitz}}]{Appleton2004}
{Appleton}, P.~N. et~al., . 2004, \apjs, 154, 147

\bibitem[{{Bacon} {et~al.}(2015){Bacon}, {Bridle}, {Brown}, {Bull}, {Camera},
  {Fender}, {Jarvis}, {Jackson}, {Kirk}, {Mann}, {McKean}, \&
  {Newman}}]{Bacon2014}
{Bacon}, D. et~al., . 2015, in proceedings of "Advancing
   Astrophysics with the Square Kilometre Array", 2015,
   \pos{PoS(AASKA14)145}

\bibitem[{{Baldry} {et~al.}(2006){Baldry}, {Balogh}, {Bower}, {Glazebrook},
  {Nichol}, {Bamford}, \& {Budavari}}]{Baldry2006}
{Baldry}, I.~K., {Balogh}, M.~L., {Bower}, R.~G., {Glazebrook}, K., {Nichol},
  R.~C., {Bamford}, S.~P., \& {Budavari}, T. 2006, \mnras, 373, 469

\bibitem[{{Baldwin} {et~al.}(1981){Baldwin}, {Phillips}, \&
  {Terlevich}}]{BPTdiagram}
{Baldwin}, J.~A., {Phillips}, M.~M., \& {Terlevich}, R. 1981, \pasp, 93, 5

\bibitem[{{Banerji} {et~al.}(2014){Banerji}, {Jouvel}, {Lin}, {McMahon},
  {Lahav}, {Castander}, {Abdalla}, {Bertin}, {Bosman}, {Carnero}, {Carrasco
  Kind}, {da Costa}, {Gerdes}, {Gschwend}, {Lima}, {Maia}, {Merson}, {Miller},
  {Ogando}, {Pellegrini}, {Reed}, {Saglia}, {Sanchez}, {Annis}, {Bernstein},
  {Bernstein}, {Bernstein}, {Capozzi}, {Childress}, {Cunha}, {Davis}, {DePoy},
  {Desai}, {Diehl}, {Doel}, {Findlay}, {Finley}, {Flaugher}, {Frieman},
  {Gaztanaga}, {Glazebrook}, {Gonzalez-Fernandez}, {Gonzalez-Solares},
  {Honscheid}, {Irwin}, {Jarvis}, {Kim}, {Koposov}, {Kuehn}, {Kupcu-Yoldas},
  {Lagattuta}, {Lewis}, {Lidman}, {Makler}, {Marriner}, {Marshall}, {Miquel},
  {Mohr}, {Neilsen}, {Peoples}, {Sako}, {Sanchez}, {Scarpine}, {Schindler},
  {Schubnell}, {Sevilla}, {Sharp}, {Soares-Santos}, {Swanson}, {Tarle},
  {Thaler}, {Tucker}, {Uddin}, {Wechsler}, {Wester}, {Yuan}, \&
  {Zuntz}}]{Banerji2014}
{Banerji}, M. et~al., . 2014, ArXiv.1407.3801

\bibitem[{{Bell}(2003)}]{Bell2003}
{Bell}, E.~F. 2003, \apj, 586, 794

\bibitem[{{Bonzini} {et~al.}(2013){Bonzini}, {Padovani}, {Mainieri},
  {Kellermann}, {Miller}, {Rosati}, {Tozzi}, \& {Vattakunnel}}]{Bonzini2013}
{Bonzini}, M., {Padovani}, P., {Mainieri}, V., {Kellermann}, K.~I., {Miller},
  N., {Rosati}, P., {Tozzi}, P., \& {Vattakunnel}, S. 2013, \mnras, 436, 3759

\bibitem[{{Bourne} {et~al.}(2011){Bourne}, {Dunne}, {Ivison}, {Maddox},
  {Dickinson}, \& {Frayer}}]{Bourne2011}
{Bourne}, N., {Dunne}, L., {Ivison}, R.~J., {Maddox}, S.~J., {Dickinson}, M.,
  \& {Frayer}, D.~T. 2011, \mnras, 410, 1155

\bibitem[{{Brown} {et~al.}(2015){Brown}, {Bacon}, {Camera}, {Harrison},
  {Joachimi}, {Metcalfe}, {Pourtsidou}, {Takahashi}, {Zuntz}, {Abdalla},
  {Bridle}, {Jarvis}, {Kitching}, {Miller}, \& {Patel}}]{Brown2014}
{Brown}, M. et~al., . 2015, in proceedings of "Advancing
   Astrophysics with the Square Kilometre Array", 2015,
   \pos{PoS(AASKA14)023}

\bibitem[{{Brown} {et~al.}(2013){Brown}, {Abdalla}, {Amara}, {Bacon}, {Battye},
  {Bell}, {Beswick}, {Birkinshaw}, {B{\"o}hm}, {Bridle}, {Browne}, {Casey},
  {Demetroullas}, {En{\ss} lin}, {Ferreira}, {Garrington}, {Grainge}, {Gray},
  {Hales}, {Harrison}, {Heavens}, {Heymans}, {Hung}, {Jackson}, {Jarvis},
  {Joachimi}, {Kay}, {Kitching}, {Leahy}, {Maartens}, {Miller}, {Muxlow},
  {Myers}, {Nichol}, {Patel}, {Pritchard}, {Raccanelli}, {Refregier},
  {Richards}, {Riseley}, {Santos}, {Scaife}, {Sch{\"a}fer}, {Schilizzi},
  {Smail}, {Starck}, {Szepietowski}, {Taylor}, {Whittaker}, {Wrigley}, \&
  {Zuntz}}]{BrownVLASS}
{Brown}, M.~L. et~al., . 2013, ArXiv.1312.5618

\bibitem[{{Burgarella} {et~al.}(2013){Burgarella}, {Buat}, {Gruppioni},
  {Cucciati}, {Heinis}, {Berta}, {B{\'e}thermin}, {Bock}, {Cooray}, {Dunlop},
  {Farrah}, {Franceschini}, {Le Floc'h}, {Lutz}, {Magnelli}, {Nordon},
  {Oliver}, {Page}, {Popesso}, {Pozzi}, {Riguccini}, {Vaccari}, \&
  {Viero}}]{Burgarella2013}
{Burgarella}, D. et~al., . 2013, \aap, 554, A70

\bibitem[{{Ciliegi} \& {Bardelli}(2015)}]{CiliegiBardelli2014}
{Ciliegi}, P. \& {Bardelli}, S. 2015, in proceedings of "Advancing
   Astrophysics with the Square Kilometre Array", 2015,
   \pos{PoS(AASKA14)150}

\bibitem[{{Cirasuolo} {et~al.}(2012){Cirasuolo}, {Afonso}, {Bender},
  {Bonifacio}, {Evans}, {Kaper}, {Oliva}, {Vanzi}, {Abreu}, {Atad-Ettedgui},
  {Babusiaux}, {Bauer}, {Best}, {Bezawada}, {Bryson}, {Cabral}, {Caputi},
  {Centrone}, {Chemla}, {Cimatti}, {Cioni}, {Clementini}, {Coelho}, {Daddi},
  {Dunlop}, {Feltzing}, {Ferguson}, {Flores}, {Fontana}, {Fynbo}, {Garilli},
  {Glauser}, {Guinouard}, {Hammer}, {Hastings}, {Hess}, {Ivison}, {Jagourel},
  {Jarvis}, {Kauffman}, {Lawrence}, {Lee}, {Li Causi}, {Lilly}, {Lorenzetti},
  {Maiolino}, {Mannucci}, {McLure}, {Minniti}, {Montgomery}, {Muschielok},
  {Nandra}, {Navarro}, {Norberg}, {Origlia}, {Padilla}, {Peacock}, {Pedicini},
  {Pentericci}, {Pragt}, {Puech}, {Randich}, {Renzini}, {Ryde}, {Rodrigues},
  {Royer}, {Saglia}, {S{\'a}nchez}, {Schnetler}, {Sobral}, {Speziali}, {Todd},
  {Tolstoy}, {Torres}, {Venema}, {Vitali}, {Wegner}, {Wells}, {Wild}, \&
  {Wright}}]{CirasuoloMOONS2012}
{Cirasuolo}, M. et~al., . 2012, in Society of Photo-Optical Instrumentation
  Engineers (SPIE) Conference Series, Vol. 8446, Society of Photo-Optical
  Instrumentation Engineers (SPIE) Conference Series

\bibitem[{{Condon}(1992)}]{Condon1992}
{Condon}, J.~J. 1992, \araa, 30, 575

\bibitem[{{Condon} {et~al.}(2012){Condon}, {Cotton}, {Fomalont}, {Kellermann},
  {Miller}, {Perley}, {Scott}, {Vernstrom}, \& {Wall}}]{Condon2012}
{Condon}, J.~J., {Cotton}, W.~D., {Fomalont}, E.~B., {Kellermann}, K.~I.,
  {Miller}, N., {Perley}, R.~A., {Scott}, D., {Vernstrom}, T., \& {Wall}, J.~V.
  2012, \apj, 758, 23

\bibitem[{{Condon} {et~al.}(1991){Condon}, {Huang}, {Yin}, \&
  {Thuan}}]{Condon1991}
{Condon}, J.~J., {Huang}, Z.-P., {Yin}, Q.~F., \& {Thuan}, T.~X. 1991, \apj,
  378, 65

\bibitem[{{Cram} {et~al.}(1998){Cram}, {Hopkins}, {Mobasher}, \&
  {Rowan-Robinson}}]{Cram1998}
{Cram}, L., {Hopkins}, A., {Mobasher}, B., \& {Rowan-Robinson}, M. 1998, \apj,
  507, 155

\bibitem[{{Daddi} {et~al.}(2007){Daddi}, {Dickinson}, {Morrison}, {Chary},
  {Cimatti}, {Elbaz}, {Frayer}, {Renzini}, {Pope}, {Alexander}, {Bauer},
  {Giavalisco}, {Huynh}, {Kurk}, \& {Mignoli}}]{Daddi2007}
{Daddi}, E. et~al., . 2007, \apj, 670, 156

\bibitem[{{de Jong} {et~al.}(2013){de Jong}, {Kuijken}, {Applegate}, {Begeman},
  {Belikov}, {Blake}, {Bout}, {Boxhoorn}, {Buddelmeijer}, {Buddendiek},
  {Cacciato}, {Capaccioli}, {Choi}, {Cordes}, {Covone}, {Dall'Ora}, {Edge},
  {Erben}, {Franse}, {Getman}, {Grado}, {Harnois-Deraps}, {Helmich},
  {Herbonnet}, {Heymans}, {Hildebrandt}, {Hoekstra}, {Huang}, {Irisarri},
  {Joachimi}, {K{\"o}hlinger}, {Kitching}, {La Barbera}, {Lacerda},
  {McFarland}, {Miller}, {Nakajima}, {Napolitano}, {Paolillo}, {Peacock},
  {Pila-Diez}, {Puddu}, {Radovich}, {Rifatto}, {Schneider}, {Schrabback},
  {Sifon}, {Sikkema}, {Simon}, {Sutherland}, {Tudorica}, {Valentijn}, {van der
  Burg}, {van Uitert}, {van Waerbeke}, {Velander}, {Kleijn}, {Viola}, \&
  {Vriend}}]{KIDS}
{de Jong}, J.~T.~A. et~al., . 2013, The Messenger, 154, 44

\bibitem[{{de Jong} {et~al.}(1985){de Jong}, {Klein}, {Wielebinski}, \&
  {Wunderlich}}]{deJong1985}
{de Jong}, T., {Klein}, U., {Wielebinski}, R., \& {Wunderlich}, E. 1985, \aap,
  147, L6

\bibitem[{{Dole} {et~al.}(2006){Dole}, {Lagache}, {Puget}, {Caputi},
  {Fern{\'a}ndez-Conde}, {Le Floc'h}, {Papovich}, {P{\'e}rez-Gonz{\'a}lez},
  {Rieke}, \& {Blaylock}}]{Dole2006}
{Dole}, H. et~al., . 2006, \aap, 451, 417

\bibitem[{{Drake} {et~al.}(2013){Drake}, {Simpson}, {Collins}, {James},
  {Baldry}, {Ouchi}, {Jarvis}, {Bonfield}, {Ono}, {Best}, {Dalton}, {Dunlop},
  {McLure}, \& {Smith}}]{Drake2013}
{Drake}, A.~B. et~al., . 2013, \mnras, 433, 796

\bibitem[{{Dye} {et~al.}(2010){Dye}, {Dunne}, {Eales}, {Smith}, {Amblard},
  {Auld}, {Baes}, {Baldry}, {Bamford}, {Blain}, {Bonfield}, {Bremer},
  {Burgarella}, {Buttiglione}, {Cameron}, {Cava}, {Clements}, {Cooray},
  {Croom}, {Dariush}, {de Zotti}, {Driver}, {Dunlop}, {Frayer}, {Fritz},
  {Gardner}, {Gomez}, {Gonzalez-Nuevo}, {Herranz}, {Hill}, {Hopkins}, {Ibar},
  {Ivison}, {Jarvis}, {Jones}, {Kelvin}, {Lagache}, {Leeuw}, {Liske},
  {Lopez-Caniego}, {Loveday}, {Maddox}, {Micha{\l}owski}, {Negrello},
  {Norberg}, {Page}, {Parkinson}, {Pascale}, {Peacock}, {Pohlen}, {Popescu},
  {Prescott}, {Rigopoulou}, {Robotham}, {Rigby}, {Rodighiero}, {Samui},
  {Scott}, {Serjeant}, {Sharp}, {Sibthorpe}, {Temi}, {Thompson}, {Tuffs},
  {Valtchanov}, {van der Werf}, {van Kampen}, \& {Verma}}]{Dye2010}
{Dye}, S. et~al., . 2010, \aap, 518, L10

\bibitem[{{Edge} {et~al.}(2013){Edge}, {Sutherland}, {Kuijken}, {Driver},
  {McMahon}, {Eales}, \& {Emerson}}]{VIKING}
{Edge}, A., {Sutherland}, W., {Kuijken}, K., {Driver}, S., {McMahon}, R.,
  {Eales}, S., \& {Emerson}, J.~P. 2013, The Messenger, 154, 32

\bibitem[{{Elbaz} {et~al.}(2011){Elbaz}, {Dickinson}, {Hwang},
  {D{\'{\i}}az-Santos}, {Magdis}, {Magnelli}, {Le Borgne}, {Galliano},
  {Pannella}, {Chanial}, {Armus}, {Charmandaris}, {Daddi}, {Aussel}, {Popesso},
  {Kartaltepe}, {Altieri}, {Valtchanov}, {Coia}, {Dannerbauer}, {Dasyra},
  {Leiton}, {Mazzarella}, {Alexander}, {Buat}, {Burgarella}, {Chary}, {Gilli},
  {Ivison}, {Juneau}, {Le Floc'h}, {Lutz}, {Morrison}, {Mullaney}, {Murphy},
  {Pope}, {Scott}, {Brodwin}, {Calzetti}, {Cesarsky}, {Charlot}, {Dole},
  {Eisenhardt}, {Ferguson}, {F{\"o}rster Schreiber}, {Frayer}, {Giavalisco},
  {Huynh}, {Koekemoer}, {Papovich}, {Reddy}, {Surace}, {Teplitz}, {Yun}, \&
  {Wilson}}]{Elbaz2011}
{Elbaz}, D. et~al., . 2011, \aap, 533, A119

\bibitem[{{Erb} {et~al.}(2003){Erb}, {Shapley}, {Steidel}, {Pettini},
  {Adelberger}, {Hunt}, {Moorwood}, \& {Cuby}}]{Erb2003}
{Erb}, D.~K., {Shapley}, A.~E., {Steidel}, C.~C., {Pettini}, M., {Adelberger},
  K.~L., {Hunt}, M.~P., {Moorwood}, A.~F.~M., \& {Cuby}, J.-G. 2003, \apj, 591,
  101

\bibitem[{{Erb} {et~al.}(2006){Erb}, {Steidel}, {Shapley}, {Pettini}, {Reddy},
  \& {Adelberger}}]{Erb2006}
{Erb}, D.~K., {Steidel}, C.~C., {Shapley}, A.~E., {Pettini}, M., {Reddy},
  N.~A., \& {Adelberger}, K.~L. 2006, \apj, 647, 128

\bibitem[{{Ferramacho} {et~al.}(2014){Ferramacho}, {Santos}, {Jarvis}, \&
  {Camera}}]{Ferramacho2014}
{Ferramacho}, L.~D., {Santos}, M.~G., {Jarvis}, M.~J., \& {Camera}, S. 2014,
  \mnras, 442, 2511

\bibitem[{{Foucaud} {et~al.}(2007){Foucaud}, {Almaini}, {Smail}, {Conselice},
  {Lane}, {Edge}, {Simpson}, {Dunlop}, {McLure}, {Cirasuolo}, {Hirst},
  {Watson}, \& {Page}}]{Foucaud2007}
{Foucaud}, S. et~al., . 2007, \mnras, 376, L20

\bibitem[{{Furusawa} {et~al.}(2008){Furusawa}, {Kosugi}, {Akiyama}, {Takata},
  {Sekiguchi}, {Tanaka}, {Iwata}, {Kajisawa}, {Yasuda}, {Doi}, {Ouchi},
  {Simpson}, {Shimasaku}, {Yamada}, {Furusawa}, {Morokuma}, {Ishida}, {Aoki},
  {Fuse}, {Imanishi}, {Iye}, {Karoji}, {Kobayashi}, {Kodama}, {Komiyama},
  {Maeda}, {Miyazaki}, {Mizumoto}, {Nakata}, {Noumaru}, {Ogasawara}, {Okamura},
  {Saito}, {Sasaki}, {Ueda}, \& {Yoshida}}]{SXDFsurvey}
{Furusawa}, H. et~al., . 2008, \apjs, 176, 1

\bibitem[{{Gilbank} {et~al.}(2010){Gilbank}, {Baldry}, {Balogh}, {Glazebrook},
  \& {Bower}}]{Gilbank2010}
{Gilbank}, D.~G., {Baldry}, I.~K., {Balogh}, M.~L., {Glazebrook}, K., \&
  {Bower}, R.~G. 2010, \mnras, 405, 2594

\bibitem[{{Gruppioni} {et~al.}(2013){Gruppioni}, {Pozzi}, {Rodighiero},
  {Delvecchio}, {Berta}, {Pozzetti}, {Zamorani}, {Andreani}, {Cimatti},
  {Ilbert}, {Le Floc'h}, {Lutz}, {Magnelli}, {Marchetti}, {Monaco}, {Nordon},
  {Oliver}, {Popesso}, {Riguccini}, {Roseboom}, {Rosario}, {Sargent},
  {Vaccari}, {Altieri}, {Aussel}, {Bongiovanni}, {Cepa}, {Daddi},
  {Dom{\'{\i}}nguez-S{\'a}nchez}, {Elbaz}, {F{\"o}rster Schreiber}, {Genzel},
  {Iribarrem}, {Magliocchetti}, {Maiolino}, {Poglitsch}, {P{\'e}rez
  Garc{\'{\i}}a}, {Sanchez-Portal}, {Sturm}, {Tacconi}, {Valtchanov},
  {Amblard}, {Arumugam}, {Bethermin}, {Bock}, {Boselli}, {Buat}, {Burgarella},
  {Castro-Rodr{\'{\i}}guez}, {Cava}, {Chanial}, {Clements}, {Conley}, {Cooray},
  {Dowell}, {Dwek}, {Eales}, {Franceschini}, {Glenn}, {Griffin},
  {Hatziminaoglou}, {Ibar}, {Isaak}, {Ivison}, {Lagache}, {Levenson}, {Lu},
  {Madden}, {Maffei}, {Mainetti}, {Nguyen}, {O'Halloran}, {Page}, {Panuzzo},
  {Papageorgiou}, {Pearson}, {P{\'e}rez-Fournon}, {Pohlen}, {Rigopoulou},
  {Rowan-Robinson}, {Schulz}, {Scott}, {Seymour}, {Shupe}, {Smith}, {Stevens},
  {Symeonidis}, {Trichas}, {Tugwell}, {Vigroux}, {Wang}, {Wright}, {Xu},
  {Zemcov}, {Bardelli}, {Carollo}, {Contini}, {Le F{\'e}vre}, {Lilly},
  {Mainieri}, {Renzini}, {Scodeggio}, \& {Zucca}}]{Gruppioni2013}
{Gruppioni}, C. et~al., . 2013, \mnras, 432, 23

\bibitem[{{Haarsma} {et~al.}(2000){Haarsma}, {Partridge}, {Windhorst}, \&
  {Richards}}]{Haarsma2000}
{Haarsma}, D.~B., {Partridge}, R.~B., {Windhorst}, R.~A., \& {Richards}, E.~A.
  2000, \apj, 544, 641

\bibitem[{{Hales}(2013)}]{HalesVLASS}
{Hales}, C.~A. 2013, ArXiv.1312.4602

\bibitem[{{Herbert} {et~al.}(2010){Herbert}, {Jarvis}, {Willott}, {McLure},
  {Mitchell}, {Rawlings}, {Hill}, \& {Dunlop}}]{Herbert2010}
{Herbert}, P.~D., {Jarvis}, M.~J., {Willott}, C.~J., {McLure}, R.~J.,
  {Mitchell}, E., {Rawlings}, S., {Hill}, G.~J., \& {Dunlop}, J.~S. 2010,
  \mnras, 406, 1841

\bibitem[{{Heywood} {et~al.}(2013){Heywood}, {Bielby}, {Hill}, {Metcalfe},
  {Rawlings}, {Shanks}, \& {Smirnov}}]{Heywood2013}
{Heywood}, I., {Bielby}, R.~M., {Hill}, M.~D., {Metcalfe}, N., {Rawlings}, S.,
  {Shanks}, T., \& {Smirnov}, O.~M. 2013, \mnras, 428, 935

\bibitem[{{Hopkins}(2004)}]{Hopkins2004}
{Hopkins}, A.~M. 2004, \apj, 615, 209

\bibitem[{{Ivison} {et~al.}(2007){Ivison}, {Greve}, {Dunlop}, {Peacock},
  {Egami}, {Smail}, {Ibar}, {van Kampen}, {Aretxaga}, {Babbedge}, {Biggs},
  {Blain}, {Chapman}, {Clements}, {Coppin}, {Farrah}, {Halpern}, {Hughes},
  {Jarvis}, {Jenness}, {Jones}, {Mortier}, {Oliver}, {Papovich},
  {P{\'e}rez-Gonz{\'a}lez}, {Pope}, {Rawlings}, {Rieke}, {Rowan-Robinson},
  {Savage}, {Scott}, {Seigar}, {Serjeant}, {Simpson}, {Stevens}, {Vaccari},
  {Wagg}, \& {Willott}}]{Ivison2007}
{Ivison}, R.~J. et~al., . 2007, \mnras, 380, 199

\bibitem[{{Ivison} {et~al.}(2010){Ivison}, {Magnelli}, {Ibar}, {Andreani},
  {Elbaz}, {Altieri}, {Amblard}, {Arumugam}, {Auld}, {Aussel}, {Babbedge},
  {Berta}, {Blain}, {Bock}, {Bongiovanni}, {Boselli}, {Buat}, {Burgarella},
  {Castro-Rodr{\'{\i}}guez}, {Cava}, {Cepa}, {Chanial}, {Cimatti}, {Cirasuolo},
  {Clements}, {Conley}, {Conversi}, {Cooray}, {Daddi}, {Dominguez}, {Dowell},
  {Dwek}, {Eales}, {Farrah}, {F{\"o}rster Schreiber}, {Fox}, {Franceschini},
  {Gear}, {Genzel}, {Glenn}, {Griffin}, {Gruppioni}, {Halpern},
  {Hatziminaoglou}, {Isaak}, {Lagache}, {Levenson}, {Lu}, {Lutz}, {Madden},
  {Maffei}, {Magdis}, {Mainetti}, {Maiolino}, {Marchetti}, {Morrison},
  {Mortier}, {Nguyen}, {Nordon}, {O'Halloran}, {Oliver}, {Omont}, {Owen},
  {Page}, {Panuzzo}, {Papageorgiou}, {Pearson}, {P{\'e}rez-Fournon}, {P{\'e}rez
  Garc{\'{\i}}a}, {Poglitsch}, {Pohlen}, {Popesso}, {Pozzi}, {Rawlings},
  {Raymond}, {Rigopoulou}, {Riguccini}, {Rizzo}, {Rodighiero}, {Roseboom},
  {Rowan-Robinson}, {Saintonge}, {Sanchez Portal}, {Santini}, {Schulz},
  {Scott}, {Seymour}, {Shao}, {Shupe}, {Smith}, {Stevens}, {Sturm},
  {Symeonidis}, {Tacconi}, {Trichas}, {Tugwell}, {Vaccari}, {Valtchanov},
  {Vieira}, {Vigroux}, {Wang}, {Ward}, {Wright}, {Xu}, \&
  {Zemcov}}]{Ivison2010b}
---. 2010, \aap, 518, L31

\bibitem[{{Jackson} \& {Rawlings}(1997)}]{JacksonRawlings1997}
{Jackson}, N. \& {Rawlings}, S. 1997, \mnras, 286, 241

\bibitem[{{Jarvis} {et~al.}(2015){Jarvis}, {Bacon}, {Blake}, {Brown},
  {Lindsay}, {Raccanelli}, {Santos}, \& {Schwarz}}]{Jarvis2014cos}
{Jarvis}, M.~J., {Bacon}, D., {Blake}, C., {Brown}, M., {Lindsay}, S.,
  {Raccanelli}, A., {Santos}, M., \& {Schwarz}, D. 2015, in proceedings of "Advancing
   Astrophysics with the Square Kilometre Array", 2015,
   \pos{PoS(AASKA14)018}

\bibitem[{{Jarvis} {et~al.}(2014){Jarvis}, {Bhatnagar}, {Bruggen}, {Ferrari},
  {Heywood}, {Hardcastle}, {Murphy}, {Taylor}, {Smirnov}, {Simpson}, {Smol{\v
  c}i{\'c}}, {Stil}, \& {van der Heyden}}]{JarvisVLASS}
{Jarvis}, M.~J. et~al., . 2014, ArXiv.1401.4018

\bibitem[{{Jarvis} {et~al.}(2013){Jarvis}, {Bonfield}, {Bruce}, {Geach},
  {McAlpine}, {McLure}, {Gonz{\'a}lez-Solares}, {Irwin}, {Lewis}, {Yoldas},
  {Andreon}, {Cross}, {Emerson}, {Dalton}, {Dunlop}, {Hodgkin}, {Le},
  {Karouzos}, {Meisenheimer}, {Oliver}, {Rawlings}, {Simpson}, {Smail},
  {Smith}, {Sullivan}, {Sutherland}, {White}, \& {Zwart}}]{Jarvis2013}
---. 2013, \mnras, 428, 1281

\bibitem[{{Jarvis} \& {Rawlings}(2004)}]{JarvisRawlings2004}
{Jarvis}, M.~J. \& {Rawlings}, S. 2004, \nar, 48, 1173

\bibitem[{{Jarvis} {et~al.}(2010){Jarvis}, {Smith}, {Bonfield}, {Hardcastle},
  {Falder}, {Stevens}, {Ivison}, {Auld}, {Baes}, {Baldry}, {Bamford}, {Bourne},
  {Buttiglione}, {Cava}, {Cooray}, {Dariush}, {de Zotti}, {Dunlop}, {Dunne},
  {Dye}, {Eales}, {Fritz}, {Hill}, {Hopwood}, {Hughes}, {Ibar}, {Jones},
  {Kelvin}, {Lawrence}, {Leeuw}, {Loveday}, {Maddox}, {Micha{\l}owski},
  {Negrello}, {Norberg}, {Pohlen}, {Prescott}, {Rigby}, {Robotham},
  {Rodighiero}, {Scott}, {Sharp}, {Temi}, {Thompson}, {van der Werf}, {van
  Kampen}, {Vlahakis}, \& {White}}]{Jarvis2010}
{Jarvis}, M.~J. et~al., . 2010, \mnras, 409, 92

\bibitem[{{Karim} {et~al.}(2011){Karim}, {Schinnerer},
  {Mart{\'{\i}}nez-Sansigre}, {Sargent}, {van der Wel}, {Rix}, {Ilbert},
  {Smol{\v c}i{\'c}}, {Carilli}, {Pannella}, {Koekemoer}, {Bell}, \&
  {Salvato}}]{Karim2011}
{Karim}, A. et~al., . 2011, \apj, 730, 61

\bibitem[{{Kewley} {et~al.}(2013){Kewley}, {Maier}, {Yabe}, {Ohta}, {Akiyama},
  {Dopita}, \& {Yuan}}]{Kewley2013}
{Kewley}, L.~J., {Maier}, C., {Yabe}, K., {Ohta}, K., {Akiyama}, M., {Dopita},
  M.~A., \& {Yuan}, T. 2013, \apjl, 774, L10

\bibitem[{{Lapi} {et~al.}(2011){Lapi}, {Gonz{\'a}lez-Nuevo}, {Fan}, {Bressan},
  {De Zotti}, {Danese}, {Negrello}, {Dunne}, {Eales}, {Maddox}, {Auld}, {Baes},
  {Bonfield}, {Buttiglione}, {Cava}, {Clements}, {Cooray}, {Dariush}, {Dye},
  {Fritz}, {Herranz}, {Hopwood}, {Ibar}, {Ivison}, {Jarvis}, {Kaviraj},
  {L{\'o}pez-Caniego}, {Massardi}, {Micha{\l}owski}, {Pascale}, {Pohlen},
  {Rigby}, {Rodighiero}, {Serjeant}, {Smith}, {Temi}, {Wardlow}, \& {van der
  Werf}}]{Lapi2011}
{Lapi}, A. et~al., . 2011, \apj, 742, 24

\bibitem[{{Madau} \& {Dickinson}(2014)}]{Madau&Dickinson2014}
{Madau}, P. \& {Dickinson}, M. 2014, \araa, 52, 415

\bibitem[{{Magnelli} {et~al.}(2014){Magnelli}, {Lutz}, {Saintonge}, {Berta},
  {Santini}, {Symeonidis}, {Altieri}, {Andreani}, {Aussel}, {B{\'e}thermin},
  {Bock}, {Bongiovanni}, {Cepa}, {Cimatti}, {Conley}, {Daddi}, {Elbaz},
  {F{\"o}rster Schreiber}, {Genzel}, {Ivison}, {Le Floc'h}, {Magdis},
  {Maiolino}, {Nordon}, {Oliver}, {Page}, {P{\'e}rez Garc{\'{\i}}a},
  {Poglitsch}, {Popesso}, {Pozzi}, {Riguccini}, {Rodighiero}, {Rosario},
  {Roseboom}, {Sanchez-Portal}, {Scott}, {Sturm}, {Tacconi}, {Valtchanov},
  {Wang}, \& {Wuyts}}]{Magnelli2014}
{Magnelli}, B. et~al., . 2014, \aap, 561, A86

\bibitem[{{Magnelli} {et~al.}(2013){Magnelli}, {Popesso}, {Berta}, {Pozzi},
  {Elbaz}, {Lutz}, {Dickinson}, {Altieri}, {Andreani}, {Aussel},
  {B{\'e}thermin}, {Bongiovanni}, {Cepa}, {Charmandaris}, {Chary}, {Cimatti},
  {Daddi}, {F{\"o}rster Schreiber}, {Genzel}, {Gruppioni}, {Harwit}, {Hwang},
  {Ivison}, {Magdis}, {Maiolino}, {Murphy}, {Nordon}, {Pannella}, {P{\'e}rez
  Garc{\'{\i}}a}, {Poglitsch}, {Rosario}, {Sanchez-Portal}, {Santini}, {Scott},
  {Sturm}, {Tacconi}, \& {Valtchanov}}]{Magnelli2013}
---. 2013, \aap, 553, A132

\bibitem[{{Makhatini} {et~al.}(2015){Makhatini}, {Smirnov}, {Jarvis}, \&
  {Heywood}}]{Makhatini2014}
{Makhatini}, S., {Smirnov}, O., {Jarvis}, M., \& {Heywood}, I. 2015, in proceedings of "Advancing
   Astrophysics with the Square Kilometre Array", 2015,
   \pos{PoS(AASKA14)081}

\bibitem[{{McAlpine} {et~al.}(2013){McAlpine}, {Jarvis}, \&
  {Bonfield}}]{McAlpine2013}
{McAlpine}, K., {Jarvis}, M.~J., \& {Bonfield}, D.~G. 2013, \mnras, 436, 1084

\bibitem[{{McAlpine} {et~al.}(2015){McAlpine}, {Prandoni}, {Jarvis}, {Seymour},
  {Padovani}, {Best}, {Simpson}, {Guidetti}, {Murphy}, {Huynh}, {Vaccari}, \&
  {White}}]{McAlpine2014}
{McAlpine}, K. et~al., . 2015, in proceedings of "Advancing
   Astrophysics with the Square Kilometre Array", 2015,
   \pos{PoS(AASKA14)083}

\bibitem[{{McCracken} {et~al.}(2012){McCracken}, {Milvang-Jensen}, {Dunlop},
  {Franx}, {Fynbo}, {Le F{\`e}vre}, {Holt}, {Caputi}, {Goranova}, {Buitrago},
  {Emerson}, {Freudling}, {Hudelot}, {L{\'o}pez-Sanjuan}, {Magnard}, {Mellier},
  {M{\o}ller}, {Nilsson}, {Sutherland}, {Tasca}, \& {Zabl}}]{McCracken2012}
{McCracken}, H.~J. et~al., . 2012, \aap, 544, A156

\bibitem[{{McKean} {et~al.}(2015){McKean}, {Jackson}, {Vegetti}, {Rybak},
  {Serjeant}, {Koopmans}, {Metcalf}, {Fassnacht}, {Marshall}, \&
  {Pandey-Pommier}}]{McKean2014}
{McKean}, J. et~al., . 2015, in proceedings of "Advancing
   Astrophysics with the Square Kilometre Array", 2015,
   \pos{PoS(AASKA14)084}

\bibitem[{{Moster} {et~al.}(2011){Moster}, {Somerville}, {Newman}, \&
  {Rix}}]{Moster2011}
{Moster}, B.~P., {Somerville}, R.~S., {Newman}, J.~A., \& {Rix}, H.-W. 2011,
  \apj, 731, 113

\bibitem[{{Murphy} {et~al.}(2015){Murphy}, {Sargent}, {Beswick}, {Dickinson},
  {Heywood}, {Hunt}, {Huynh}, {Jarvis}, {Karim}, {Krause}, {Prandoni},
  {Seymour}, {Schinnerer}, {Tabatabei}, \& {Wagg}}]{Murphy2014}
{Murphy}, E. et~al., . 2015, in proceedings of "Advancing
   Astrophysics with the Square Kilometre Array", 2015,
   \pos{PoS(AASKA14)085}

\bibitem[{{Murphy}(2009)}]{Murphy2009}
{Murphy}, E.~J. 2009, \apj, 706, 482

\bibitem[{{Muxlow} {et~al.}(2005){Muxlow}, {Richards}, {Garrington},
  {Wilkinson}, {Anderson}, {Richards}, {Axon}, {Fomalont}, {Kellermann},
  {Partridge}, \& {Windhorst}}]{Muxlow2005}
{Muxlow}, T.~W.~B. et~al., . 2005, \mnras, 358, 1159

\bibitem[{{Noeske} {et~al.}(2007){Noeske}, {Weiner}, {Faber}, {Papovich},
  {Koo}, {Somerville}, {Bundy}, {Conselice}, {Newman}, {Schiminovich}, {Le
  Floc'h}, {Coil}, {Rieke}, {Lotz}, {Primack}, {Barmby}, {Cooper}, {Davis},
  {Ellis}, {Fazio}, {Guhathakurta}, {Huang}, {Kassin}, {Martin}, {Phillips},
  {Rich}, {Small}, {Willmer}, \& {Wilson}}]{Noeske2007}
{Noeske}, K.~G. et~al., . 2007, \apjl, 660, L43

\bibitem[{{Padovani} {et~al.}(2009){Padovani}, {Mainieri}, {Tozzi},
  {Kellermann}, {Fomalont}, {Miller}, {Rosati}, \& {Shaver}}]{Padovani2009}
{Padovani}, P., {Mainieri}, V., {Tozzi}, P., {Kellermann}, K.~I., {Fomalont},
  E.~B., {Miller}, N., {Rosati}, P., \& {Shaver}, P. 2009, \apj, 694, 235

\bibitem[{{Peng} {et~al.}(2010){Peng}, {Lilly}, {Kova{\v c}}, {Bolzonella},
  {Pozzetti}, {Renzini}, {Zamorani}, {Ilbert}, {Knobel}, {Iovino}, {Maier},
  {Cucciati}, {Tasca}, {Carollo}, {Silverman}, {Kampczyk}, {de Ravel},
  {Sanders}, {Scoville}, {Contini}, {Mainieri}, {Scodeggio}, {Kneib}, {Le
  F{\`e}vre}, {Bardelli}, {Bongiorno}, {Caputi}, {Coppa}, {de la Torre},
  {Franzetti}, {Garilli}, {Lamareille}, {Le Borgne}, {Le Brun}, {Mignoli},
  {Perez Montero}, {Pello}, {Ricciardelli}, {Tanaka}, {Tresse}, {Vergani},
  {Welikala}, {Zucca}, {Oesch}, {Abbas}, {Barnes}, {Bordoloi}, {Bottini},
  {Cappi}, {Cassata}, {Cimatti}, {Fumana}, {Hasinger}, {Koekemoer},
  {Leauthaud}, {Maccagni}, {Marinoni}, {McCracken}, {Memeo}, {Meneux}, {Nair},
  {Porciani}, {Presotto}, \& {Scaramella}}]{Peng2010}
{Peng}, Y.-j. et~al., . 2010, \apj, 721, 193

\bibitem[{{Rau} {et~al.}(2014){Rau}, {Bhatnagar}, \& {Owen}}]{Rau2014}
{Rau}, U., {Bhatnagar}, S., \& {Owen}, F.~N. 2014, ArXiv.1403.5242

\bibitem[{{Rodighiero} {et~al.}(2014){Rodighiero}, {Renzini}, {Daddi},
  {Baronchelli}, {Berta}, {Cresci}, {Franceschini}, {Gruppioni}, {Lutz},
  {Mancini}, {Santini}, {Zamorani}, {Silverman}, {Kashino}, {Andreani},
  {Cimatti}, {S{\'a}nchez}, {Le Floch}, {Magnelli}, {Popesso}, \&
  {Pozzi}}]{Rodighiero2014}
{Rodighiero}, G. et~al., . 2014, \mnras, 443, 19

\bibitem[{{Scoville} {et~al.}(2013){Scoville}, {Arnouts}, {Aussel}, {Benson},
  {Bongiorno}, {Bundy}, {Calvo}, {Capak}, {Carollo}, {Civano}, {Dunlop},
  {Elvis}, {Faisst}, {Finoguenov}, {Fu}, {Giavalisco}, {Guo}, {Ilbert},
  {Iovino}, {Kajisawa}, {Kartaltepe}, {Leauthaud}, {Le F{\`e}vre}, {LeFloch},
  {Lilly}, {Liu}, {Manohar}, {Massey}, {Masters}, {McCracken}, {Mobasher},
  {Peng}, {Renzini}, {Rhodes}, {Salvato}, {Sanders}, {Sarvestani}, {Scarlata},
  {Schinnerer}, {Sheth}, {Shopbell}, {Smol{\v c}i{\'c}}, {Taniguchi}, {Taylor},
  {White}, \& {Yan}}]{Scoville2013}
{Scoville}, N. et~al., . 2013, \apjs, 206, 3

\bibitem[{{Scoville} {et~al.}(2007){Scoville}, {Aussel}, {Brusa}, {Capak},
  {Carollo}, {Elvis}, {Giavalisco}, {Guzzo}, {Hasinger}, {Impey}, {Kneib},
  {LeFevre}, {Lilly}, {Mobasher}, {Renzini}, {Rich}, {Sanders}, {Schinnerer},
  {Schminovich}, {Shopbell}, {Taniguchi}, \& {Tyson}}]{Scoville2007}
---. 2007, \apjs, 172, 1

\bibitem[{{Seymour} {et~al.}(2008){Seymour}, {Dwelly}, {Moss}, {McHardy},
  {Zoghbi}, {Rieke}, {Page}, {Hopkins}, \& {Loaring}}]{Seymour2008}
{Seymour}, N., {Dwelly}, T., {Moss}, D., {McHardy}, I., {Zoghbi}, A., {Rieke},
  G., {Page}, M., {Hopkins}, A., \& {Loaring}, N. 2008, \mnras, 386, 1695

\bibitem[{{Simpson} {et~al.}(2006){Simpson}, {Mart{\'{\i}}nez-Sansigre},
  {Rawlings}, {Ivison}, {Akiyama}, {Sekiguchi}, {Takata}, {Ueda}, \&
  {Watson}}]{Simpson2006}
{Simpson}, C., {Mart{\'{\i}}nez-Sansigre}, A., {Rawlings}, S., {Ivison}, R.,
  {Akiyama}, M., {Sekiguchi}, K., {Takata}, T., {Ueda}, Y., \& {Watson}, M.
  2006, \mnras, 372, 741

\bibitem[{{Smith} {et~al.}(2014){Smith}, {Jarvis}, {Hardcastle}, {Vaccari},
  {Bourne}, {Dunne}, {Ibar}, {Maddox}, {Prescott}, {Vlahakis}, {Eales},
  {Maddox}, {Smith}, {Valiante}, \& {de Zotti}}]{Smith2014}
{Smith}, D.~J.~B. et~al., . 2014, \mnras, 445, 2232

\bibitem[{{Smol{\v c}i{\'c}} {et~al.}(2015){Smol{\v c}i{\'c}}, {Padovani},
  {Delhaize}, {Prandoni}, {Seymour}, {Jarvis}, {Afonso}, {Magliocchetti},
  {Huynh}, {Vaccari}, \& {Karim}}]{Smolcic2014}
{Smol{\v c}i{\'c}}, E. et~al., . 2015,in proceedings of "Advancing
   Astrophysics with the Square Kilometre Array", 2015,
   \pos{PoS(AASKA14)069}

\bibitem[{{Smol{\v c}i{\'c}} {et~al.}(2009{\natexlab{a}}){Smol{\v c}i{\'c}},
  {Schinnerer}, {Zamorani}, {Bell}, {Bondi}, {Carilli}, {Ciliegi}, {Mobasher},
  {Paglione}, {Scodeggio}, \& {Scoville}}]{Smolcic2009sf}
{Smol{\v c}i{\'c}}, V. et~al., . 2009{\natexlab{a}}, \apj, 690, 610

\bibitem[{{Smol{\v c}i{\'c}} {et~al.}(2009{\natexlab{b}}){Smol{\v c}i{\'c}},
  {Zamorani}, {Schinnerer}, {Bardelli}, {Bondi}, {B{\^i}rzan}, {Carilli},
  {Ciliegi}, {Elvis}, {Impey}, {Koekemoer}, {Merloni}, {Paglione}, {Salvato},
  {Scodeggio}, {Scoville}, \& {Trump}}]{Smolcic2009}
---. 2009{\natexlab{b}}, \apj, 696, 24

\bibitem[{{Sobral} {et~al.}(2012){Sobral}, {Best}, {Matsuda}, {Smail}, {Geach},
  \& {Cirasuolo}}]{Sobral2012}
{Sobral}, D., {Best}, P.~N., {Matsuda}, Y., {Smail}, I., {Geach}, J.~E., \&
  {Cirasuolo}, M. 2012, \mnras, 420, 1926

\bibitem[{{Takada} {et~al.}(2014){Takada}, {Ellis}, {Chiba}, {Greene},
  {Aihara}, {Arimoto}, {Bundy}, {Cohen}, {Dor{\'e}}, {Graves}, {Gunn},
  {Heckman}, {Hirata}, {Ho}, {Kneib}, {F{\`e}vre}, {Lin}, {More}, {Murayama},
  {Nagao}, {Ouchi}, {Seiffert}, {Silverman}, {Sodr{\'e}}, {Spergel}, {Strauss},
  {Sugai}, {Suto}, {Takami}, \& {Wyse}}]{TakadaPFS2014}
{Takada}, M. et~al., . 2014, \pasj, 66, 1

\bibitem[{{Takeuchi} {et~al.}(2005){Takeuchi}, {Buat}, \&
  {Burgarella}}]{Takeuchi2005}
{Takeuchi}, T.~T., {Buat}, V., \& {Burgarella}, D. 2005, \aap, 440, L17

\bibitem[{{Vaccari} {et~al.}(2010){Vaccari}, {Marchetti}, {Franceschini},
  {Altieri}, {Amblard}, {Arumugam}, {Auld}, {Aussel}, {Babbedge}, {Blain},
  {Bock}, {Boselli}, {Buat}, {Burgarella}, {Castro-Rodr{\'{\i}}guez}, {Cava},
  {Chanial}, {Clements}, {Conley}, {Conversi}, {Cooray}, {Dowell}, {Dwek},
  {Dye}, {Eales}, {Elbaz}, {Farrah}, {Fox}, {Gear}, {Glenn}, {Gonz{\'a}lez
  Solares}, {Griffin}, {Halpern}, {Hatziminaoglou}, {Huang}, {Ibar}, {Isaak},
  {Ivison}, {Lagache}, {Levenson}, {Lu}, {Madden}, {Maffei}, {Mainetti},
  {Mortier}, {Nguyen}, {O'Halloran}, {Oliver}, {Omont}, {Page}, {Panuzzo},
  {Papageorgiou}, {Pearson}, {P{\'e}rez-Fournon}, {Pohlen}, {Rawlings},
  {Raymond}, {Rigopoulou}, {Rizzo}, {Rodighiero}, {Roseboom}, {Rowan-Robinson},
  {S{\'a}nchez Portal}, {Schulz}, {Scott}, {Seymour}, {Shupe}, {Smith},
  {Stevens}, {Symeonidis}, {Trichas}, {Tugwell}, {Valiante}, {Valtchanov},
  {Vigroux}, {Wang}, {Ward}, {Wright}, {Xu}, \& {Zemcov}}]{Vaccari2010}
{Vaccari}, M. et~al., . 2010, \aap, 518, L20

\bibitem[{{Vernstrom} {et~al.}(2014){Vernstrom}, {Scott}, {Wall}, {Condon},
  {Cotton}, {Fomalont}, {Kellermann}, {Miller}, \& {Perley}}]{Vernstrom2014}
{Vernstrom}, T., {Scott}, D., {Wall}, J.~V., {Condon}, J.~J., {Cotton}, W.~D.,
  {Fomalont}, E.~B., {Kellermann}, K.~I., {Miller}, N., \& {Perley}, R.~A.
  2014, \mnras, 440, 2791

\bibitem[{{Whitaker} {et~al.}(2012){Whitaker}, {van Dokkum}, {Brammer}, \&
  {Franx}}]{Whitaker2012}
{Whitaker}, K.~E., {van Dokkum}, P.~G., {Brammer}, G., \& {Franx}, M. 2012,
  \apjl, 754, L29

\bibitem[{{White} {et~al.}(2014){White}, {Jarvis}, {H{\"a}u{\ss}ler}, \&
  {Maddox}}]{White2014}
{White}, S.~V., {Jarvis}, M.~J., {H{\"a}u{\ss}ler}, B., \& {Maddox}, N. 2014,
  ArXiv.1410.3892

\bibitem[{{Willott} {et~al.}(2013){Willott}, {McLure}, {Hibon}, {Bielby},
  {McCracken}, {Kneib}, {Ilbert}, {Bonfield}, {Bruce}, \&
  {Jarvis}}]{Willott2013}
{Willott}, C.~J. et~al., . 2013, \aj, 145, 4

\bibitem[{{Wilman} {et~al.}(2010){Wilman}, {Jarvis}, {Mauch}, {Rawlings}, \&
  {Hickey}}]{Wilman2010}
{Wilman}, R.~J., {Jarvis}, M.~J., {Mauch}, T., {Rawlings}, S., \& {Hickey}, S.
  2010, \mnras, 405, 447

\bibitem[{{Wilman} {et~al.}(2008){Wilman}, {Miller}, {Jarvis}, {Mauch},
  {Levrier}, {Abdalla}, {Rawlings}, {Kl{\"o}ckner}, {Obreschkow}, {Olteanu}, \&
  {Young}}]{Wilman2008}
{Wilman}, R.~J. et~al., . 2008, \mnras, 388, 1335

\bibitem[{{Wright} {et~al.}(2010){Wright}, {Eisenhardt}, {Mainzer}, {Ressler},
  {Cutri}, {Jarrett}, {Kirkpatrick}, {Padgett}, {McMillan}, {Skrutskie},
  {Stanford}, {Cohen}, {Walker}, {Mather}, {Leisawitz}, {Gautier}, {McLean},
  {Benford}, {Lonsdale}, {Blain}, {Mendez}, {Irace}, {Duval}, {Liu}, {Royer},
  {Heinrichsen}, {Howard}, {Shannon}, {Kendall}, {Walsh}, {Larsen}, {Cardon},
  {Schick}, {Schwalm}, {Abid}, {Fabinsky}, {Naes}, \& {Tsai}}]{WISE}
{Wright}, E.~L. et~al., . 2010, \aj, 140, 1868

\bibitem[{{Yun} {et~al.}(2001){Yun}, {Reddy}, \& {Condon}}]{Yun2001}
{Yun}, M.~S., {Reddy}, N.~A., \& {Condon}, J.~J. 2001, \apj, 554, 803

\bibitem[{{Zwart} {et~al.}(2014){Zwart}, {Jarvis}, {Deane}, {Bonfield},
  {Knowles}, {Madhanpall}, {Rahmani}, \& {Smith}}]{Zwart2014}
{Zwart}, J.~T.~L., {Jarvis}, M.~J., {Deane}, R.~P., {Bonfield}, D.~G.,
  {Knowles}, K., {Madhanpall}, N., {Rahmani}, H., \& {Smith}, D.~J.~B. 2014,
  \mnras, 439, 1459

\end{thebibliography}

\end{document}